\documentclass[12pt]{article}
\usepackage{enumerate}
\usepackage{xr,natbib}
\usepackage{url} 
\makeatletter
\newcommand*{\addFileDependency}[1]{
\typeout{(#1)}
\@addtofilelist{#1}
\IfFileExists{#1}{}{\typeout{No file #1.}}
}\makeatother

\newcommand*{\myexternaldocument}[1]{%
\externaldocument{#1}%
\addFileDependency{#1.tex}%
\addFileDependency{#1.aux}%
}

\myexternaldocument{ms-Supp}
\newcommand{\blind}{1}

\usepackage{longtable}
\usepackage{bicaption}

\usepackage{bbm}
\usepackage{amsmath,amssymb,amsthm}
\usepackage{mathtools}
\usepackage[dvips]{graphicx,color}
\usepackage{adjustbox}
\usepackage{mathrsfs}
\usepackage{bm, xcolor}
\usepackage{mathrsfs}
\usepackage{verbatim}
\usepackage{longtable}
\usepackage{threeparttable}
\usepackage{booktabs,multirow}
\usepackage{color}
\usepackage{subcaption}
\usepackage{algorithm,caption}
\usepackage{algpseudocode}
\usepackage[bookmarks=false,hypertexnames=false]{hyperref}\hypersetup{colorlinks=true, linkcolor=blue, filecolor=gray, urlcolor=blue, citecolor=black}

\makeatletter
\def\singlespace{\def\baselinestretch{1}\@normalsize}

\newtheorem{lemma}{Lemma}
\newtheorem{theorem}{Theorem}
\newtheorem{remark}{Remark}
\newtheorem{corollary}{Corollary}
\@addtoreset{equation}{section}

\renewcommand{\hat}{\widehat}
\def\singlespace{\def\baselinestretch{1}\@normalsize}

\makeatother

\newcommand{\bX}{\bm{X}}

\newcommand{\bmu}{\bm{\mu}}

\newcommand{\bR}{\bm{R}}

\newcommand{\balpha}{\bm{\alpha}}
\newcommand{\btheta}{\bm{\theta}}

\newcommand{\bdelta}{\bm{\delta}}

\newcommand{\bw}{\bm{w}}

\newcommand{\br}{\bm{r}}

\newcommand{\bSigma}{\bm{\Sigma}}

\usepackage{enumitem}

\begin{document}

\def\spacingset#1{\renewcommand{\baselinestretch}%
{#1}\small\normalsize} \spacingset{1}


\title{Cost-aware Portfolios in a Large Universe of Assets}
\if1\blind
{
  \title{\bf Cost-aware Portfolios in a Large Universe of Assets}
  \author{Qingliang Fan$^1$, Marcelo C. Medeiros$^2$
    , Hanming Yang$^3$ and Songshan Yang$^3$\thanks{All the authors made equal contributions to this work and are listed alphabetically.
  The authors gratefully acknowledge \textit{please remember to list all relevant funding sources in the unblinded version}}\hspace{.2cm}\\
    $^1$The Chinese University of Hong Kong\\
   $^2$The University of Illinois at Urbana Champaign \\
   $^3$Renmin University of China
  }
\date{}
  \maketitle
} \fi

\if0\blind
{
  \title{\bf Cost-aware Portfolios in a Large Universe of Assets}
  \author{}
  \date{}
  \maketitle
} \fi

\bigskip
\begin{abstract}
This paper proposes a finite-horizon (multi-period) mean-variance portfolio estimator, in which rebalancing decisions are based on current information about asset returns and transaction costs. The novelty of this study stems from integrating transaction costs into the decision process within a high-dimensional portfolio setting, where the number of assets exceeds the sample size. We define the optimal cost-aware portfolio and propose novel models for its construction and rebalancing. Our approach incorporates a nonconvex penalty and explicitly accounts for both proportional and quadratic transaction costs. We establish that the estimators derived from the proposed construction and rebalancing models, as well as the corresponding in-sample and out-of-sample Sharpe ratio estimators, consistently converge to those of the optimal cost-aware portfolio.
Monte Carlo simulations and empirical studies using S\&P 500 and Russell 2000 stocks show the satisfactory performance of the proposed portfolio and highlight the importance of incorporating transaction costs during rebalancing.\\

\end{abstract}

\noindent%

\vspace{0.3cm}

\noindent%
{\it Acknowledgments:}  The authors are very thankful to seminar participants at the University of Wisconsin-Madison, Queen Mary University of London, Bristol University, University of California at Davis, and Vanderbilt University.

\vfill

\newpage
\spacingset{1.75} 

\section{Introduction}\label{sec:intro}

Portfolio management attracts great interest in statistics, financial econometrics, and quantitative finance. The mean-variance portfolio allocation model proposed by \citet{markowitz1952portfolio} stands as a cornerstone of modern portfolio theory. The model's practicality in terms of high dimensionality and trading frictions (e.g., transaction costs) is further studied in this paper. 

The classic Markowitz mean-variance formulation considers the following optimization problem in a single period:
\begin{equation}
\label{port_opt_exp}
\begin{aligned}
\bw^*=\arg\min_{\bw} \bw^{\top} \bSigma \bw  - \gamma \bw^{\top}\bmu,
\end{aligned}
\end{equation}
where $\bw$ is the weight vector of the assets in the portfolio, $\gamma$ is the inverse of the risk aversion parameter, $\bmu = (\mu_1, \cdots, \mu_p)^{\top}$ is the mean vector of the excess returns of $p$ assets and $\bSigma$ is the covariance matrix of the excess returns. Here we assume that $\sum_{i=1}^{p} w_i = 1$, so the portfolio is fully-invested on the pool of $p$ risky assets. \cite{multiperiod} extends the \cite{markowitz1952portfolio}model to consider discrete multi-period portfolio allocation. This approach has an elegant analytical solution that depends only on the mean and covariance matrix of the excess returns. The solution to \eqref{port_opt_exp} is $\bw^{*}=c_1 \bSigma^{-1}\bmu - c_2 \bSigma^{-1}\mathbf{1}$, where $c_1$ and $c_2$ depend only on $\bSigma$ and $\bmu$. However, in modern portfolio management, portfolios commonly exhibit high-dimensional characteristics, introducing new challenges to the traditional framework. Notably, the inverse of the sample covariance matrix is not well defined when the number of assets is larger than the sample size.

Besides the difficulties of covariance matrix estimation in high-dimensional portfolio allocation, transaction costs also make the classical Markowitz mean-variance model sub-optimal. In most empirical studies, transaction costs are retrospectively incorporated by analyzing a portfolio strategy's performance under given-sized transaction costs. However, in financial practice, the costs of the portfolio are considered during the decision-making process, meaning the costs are an integral part of the rebalancing. Transaction costs are generally assumed to be proportional to the trading amount for minor trades with negligible market impact.  For larger trades, the literature usually assumes that they have a nonlinear impact on the market price, resulting in quadratic transaction costs \citep{olivares2018robust}. \cite{Hautsch_2019} illustrated that when incorporating transaction costs into the optimization problem, quadratic transaction costs can be understood as causing the covariance matrix to contract towards a diagonal matrix and a mean shift proportionate to current holdings. Proportional transaction costs scale with the sum of the absolute values of rebalancing amounts, penalizing turnover more heavily and thereby reducing rebalancing volume and frequency. Some studies have provided empirical evidence that strategies incorporating transaction fees consistently demonstrate superior performance compared to the minimum variance benchmark methods \citep{Cardinality_Constraints, Hautsch_2019}. 

\subsection{Main Takeaways}

This paper considers the problem faced by a real mutual fund or sophisticated investor. Transaction costs are an integral part of portfolio management and are passed on to investors, who receive returns net of such costs. We propose an innovative method for constrained portfolio selection tailored for high-dimensional portfolios, taking into account two different types of transaction costs. Moreover, our approach allows for adaptive weight adjustments in multi-stage scenarios. Specifically, at each decision point, the portfolio weights are rebalanced in a data-driven manner based on the current holdings and the return information from the previous stage. In other words, we construct the optimal portfolio that considers the cost of rebalancing, which provides guidance to cost-aware investors.

The main contributions of our work are threefold. Firstly, we propose cost-aware optimal portfolio estimators for new construction and subsequent rebalancing in a high-dimensional framework. To our knowledge, such a portfolio estimator in high dimensions appears to be novel. Also, the nonconvex penalization employed in our approach can improve the estimation accuracy compared to the $\ell_1$ penalization method. We establish theoretical properties for the proposed portfolio in both the construction and rebalancing stages under each cost function. We show that, under both cost functions, the estimators obtained in the construction and rebalancing stages—as well as the corresponding in-sample and out-of-sample Sharpe ratio estimators—consistently converge to those of the optimal cost-aware portfolio. Secondly, we adopt the local linear approximation (LLA) algorithm \citep{zou2008one} to solve the nonconvex optimization problem and develop the theoretical foundations of the proposed portfolio with LLA algorithm. Thirdly, we introduce data-dependent penalization for investors facing a large universe of assets in each period while considering transaction costs. We demonstrate the superior performance of the proposed portfolio through extensive simulation studies and empirical evaluations. In particular, we compare it against a wide range of existing portfolio construction methods. For the real data analysis, we use stock return data from the S\&P 500 and Russell 2000 indices over the period 2017–2020.

\subsection{Brief Literature Review}
Our paper connects to the literature on covariance learning and high-dimensional portfolio optimization.

\citet{FAN2015367} studied a factor-based risk estimator under many assets and introduced a high-confidence level upper bound to assess the estimation. \citet{LAM2018226} proposed a nonparametrically eigenvalue-regularized integrated covariance matrix estimator. \citet{fan2019robust} constructed a robust covariance estimation under the approximate factor model with observed factors. \citet{HAFNER2020431} developed a Kronecker product model for the covariance matrix estimation and established central limit theorems for such estimator. \citet{ren2015asymptotic}, \citet{fan2016innovated} and \citet{ren2019tuning} conducted studies on the estimation and statistical inference of sparse precision matrix under the assumption of conditional normal distribution. Nevertheless, practical challenges arise in verifying the normality of high-dimensional data. \citet{WANG2020118} introduced a refitted cross-validation estimation method for a high-dimensional precision matrix utilizing its Cholesky decomposition. Other recent works include but are not limited to \citet{doi:10.1080/07350015.2017.1345683,CAI2020482,SO2020,DING2021502}.

To enhance out-of-sample performance, \citet{luo2024portfolio} proposed an iterative optimization framework based on almost second-degree stochastic dominance (ASSD), which improves convergence efficiency and empirically yields higher out-of-sample returns compared to traditional dominance and mean-variance approaches.

The estimators in the works above converge to the true covariance matrix or precision matrix under certain regularity conditions. However, whether the derived portfolio allocation provides the optimal strategy with cost consideration is still unclear.

Another strand of literature considers various constraints on large portfolio optimization. \citet{jagannathan2003risk} proposed no-short-selling constraints to reduce the portfolio risks. Inspired by \citet{jagannathan2003risk}, \citet{demiguel2009generalized} imposed the norm constraints to the portfolios. In particular, they developed a unified framework for portfolio optimization by constraining the norm of the portfolio weights, which effectively mitigates estimation error and encompasses several existing strategies such as shrinkage-based estimators and the 1/N portfolio.

\cite{fan2012vast} proposed a gross-exposure constraint method to modify the Markowitz mean-variance optimization problem in a single period to reduce the sensitivity of the resulting allocation to input vectors. \citet{Cardinality_Constraints} showed that under mild conditions, solving the $\ell_1$ constrained problem yields the same expected utility as solving the $\ell_0$ constrained problem, enabling effective implementation of cardinality constraints through $\ell_1$ constraints. \citet{zou2006adaptive,zhao2006model,gai2013model} suggested that $\ell_1$ regularization methods demonstrate the support recovery property under restrictive regularity conditions while being prone to bias for large coefficients. In contrast, nonconvex penalization methods such as SCAD \citep{fan2001scadli} and MCP \citep{zhang2010nearly} exhibit favorable theoretical properties without relying on restrictive conditions and effectively mitigate the estimation bias associated with Lasso.

In addition, \citet{ban2018machine} proposed performance-based regularization and cross-validation techniques for portfolio optimization, casting the problem into a robust optimization framework and showing that their method outperforms traditional and regularized approaches across several Fama-French datasets.

On the other hand, the literature on transaction costs and portfolio optimization in large dimensions is significantly less abundant. Most of the papers in Mathematical Finance and Operations Research focus on low-dimensional settings. See, for example, \citet{lobo2007portfolio} and \citet{olivares2018robust}. One exception is the important work by \citet{ledoit2025markowitz}, where the authors also consider transaction costs at the optimization stage when constructing mean-variance portfolios. However, their paper focuses only on the empirical part and provides no theoretical guidance. Furthermore, the authors do not consider extra regularization terms to discipline the portfolio in large dimensions. In our opinion, our paper complements the analysis in \citet{ledoit2025markowitz}.

From an algorithmic perspective, \citet{best2005algorithm} propose an efficient method for portfolio optimization with transaction costs, which reduces a high-dimensional constrained problem into a sequence of lower-dimensional subproblems by implicitly incorporating transaction cost effects. Similarly, \citet{perold1984large} provides a computationally efficient algorithm for large-scale mean-variance portfolio optimization with transaction costs and factor-based models, enabling parametric solutions under practical resource constraints.

In the dynamic setting, \citet{brown2011dynamic} study multi-period portfolio optimization with transaction costs and propose heuristic strategies that perform close to optimal, supported by dual-based upper bounds derived from information relaxation.
\subsection{Organization of the Paper}

The remainder of the article is organized as follows. Section \ref{sec:meth} proposes the constrained nonconvex penalization model for high-dimensional portfolios in both construction and rebalacing stages with consideration of transaction cost. We also introduce the LLA to solve the proposed model. 
In Section \ref{sec:numerical studies}, we conduct a comprehensive numerical study to examine the performance of the proposed method. Empirical analysis is included in Section \ref{sec:realdata}. We summarize our contribution in Section \ref{sec:conc}.  

\noindent\textbf{Notations}: We use the standard convention that constants $c_1$, $c_2$, etc., denote universal positive constants whose values may vary from line to line. We write $f(n) = O(g(n))$ if $f(n) \leq c g(n)$ for some constant $c > 0$. For a matrix $\boldsymbol{M}$, we let $|||\boldsymbol{M}|||_{\infty}$ denote the $L_{\infty}$ operator norm, $\|\boldsymbol{M}\|_{\max} = \max |m_{ij}|$ denote the entrywise $L_{\infty}$ norm, and $|||\boldsymbol{M}|||_2$ denote the spectral norm. The minimal and maximal eigenvalues of $\boldsymbol{M}$ are denoted by $\lambda_{\min}(\boldsymbol{M})$ and $\lambda_{\max}(\boldsymbol{M})$, respectively. For a vector $\boldsymbol{v} = (v_j)_{j=1}^{p} \in \mathbb{R}^{p}$, we use $\|\boldsymbol{v}\|_k$ ($k=1,2$) for its $L_k$ norm, $\|\boldsymbol{v}\|_0$ for the number of nonzero entries, and $\|\boldsymbol{v}\|_{\infty} = \max |v_j|$ for the $L_{\infty}$ norm.

Given index sets $\mathcal{A}, \mathcal{B} \subset \{1, \ldots, p\}$, we write $\boldsymbol{M}_{\mathcal{A},\mathcal{B}}$ for the submatrix of $\boldsymbol{M}$ with rows indexed by $\mathcal{A}$ and columns by $\mathcal{B}$, and $\boldsymbol{v}_{\mathcal{A}}$ for the subvector of $\boldsymbol{v}$ with entries in $\mathcal{A}$. The cardinality of $\mathcal{A}$ is denoted $|\mathcal{A}|$. The symbol $\odot$ represents element-wise multiplication. For a function $h: \mathbb{R}^p \rightarrow \mathbb{R}$, $\nabla h$ denotes its gradient or subgradient.
To simplify notation, when referring to time-specific quantities such as $\boldsymbol{\mu}_t$ and $\boldsymbol{\Sigma}_t$, we write $\boldsymbol{\mu}_{\mathcal{A}_t}$ and $\boldsymbol{\Sigma}_{\mathcal{A}_t,\mathcal{A}_t}$ in place of the more precise $\boldsymbol{\mu}_{t,\mathcal{A}_t}$ and $\boldsymbol{\Sigma}_{t,\mathcal{A}_t,\mathcal{A}_t}$ whenever the context is clear.

\section{Methodology}\label{sec:Method}
\label{sec:meth}

\subsection{Preliminary}
\label{Preliminary}

To contextualize our proposed approach, we review three foundational models in the mean-variance portfolio optimization literature. These models serve as natural precursors to our method and are also employed as benchmark comparisons in our simulation studies.

\begin{itemize}
\item[1.] \textbf{The classical Markowitz mean-variance model}. 
Originally introduced by \citet{markowitz1952portfolio}, this model aims to balance risk and return by minimizing portfolio variance for a given level of expected return. It is formulated as:
\begin{equation}
\label{baseline1}
\begin{aligned}
\min_{\bw}\ \bw^{\top} \hat{\bSigma} \bw - \gamma \bw^{\top}\hat{\bmu},\quad \text{subject to } \sum_{i=1}^{p} w_{i} = 1,
\end{aligned}
\end{equation}
where $w_{i}$ denotes the $i$th element of the portfolio weight vector $\bw$, $\hat{\bSigma}$ is the estimated covariance matrix, and $\hat{\bmu}$ is the vector of estimated expected returns. Despite its theoretical elegance, the Markowitz model often suffers from poor empirical performance in high-dimensional settings due to estimation errors in the covariance matrix.

\item[2.] \textbf{The penalized mean-variance model with gross-exposure constraints}. 
\citet{fan2012vast} propose a regularized version of the classical mean-variance framework by incorporating an $\ell_1$ penalty on the portfolio weights to control gross exposure and improve stability in high-dimensional settings. The optimization problem is:
\begin{equation}
\label{baseline2}
\begin{aligned}
\min_{\bw}\ \bw^{\top} \hat{\bSigma} \bw - \gamma \bw^{\top}\hat{\bmu} + \lambda \Vert \bw\Vert_1,\quad \text{subject to } \sum_{i=1}^{p} w_{i} = 1,
\end{aligned}
\end{equation}
where $\lambda > 0$ is a tuning parameter that regulates the portfolio's gross exposure. This formulation mitigates the accumulation of estimation errors in large portfolios and can outperform no-short-sale constraints by allowing limited short positions.

\item[3.] \textbf{The mean-variance model with transaction costs}. 
\citet{Hautsch_2019} extend the mean-variance paradigm to account for transaction costs explicitly. By penalizing turnover directly, their model promotes more stable allocations while implicitly regularizing the covariance matrix. The corresponding optimization problem is:
\begin{equation}
\label{baseline3}
\begin{aligned}
\min_{\bw}\ \bw^{\top} \hat{\bSigma}\bw - \gamma \bw^{\top}\hat{\bmu} + \mathbf{C}(\bw),\quad \text{subject to } \sum_{i=1}^{p} w_{i} = 1,
\end{aligned}
\end{equation}
where $\mathbf{C}(\bw)$ denotes the transaction cost function, often proportional or quadratic in the change of portfolio weights. This model demonstrates that incorporating transaction costs ex ante leads to empirically superior and more stable portfolios, especially under model uncertainty.

\end{itemize}

Each of these models contributes essential insights: the classical mean-variance framework provides the theoretical foundation, the penalized model addresses high-dimensional estimation risk, and the cost-aware model introduces practical considerations arising in real-world trading environments. Our proposed approach, cost-aware portfolio estimator (CAPE), builds upon these developments by integrating both regularization and dynamic transaction cost modeling in a unified framework, tailored for large-scale portfolio optimization.
\subsection{Cost-aware Portfolio Estimator}\label{section 2.1}
\begin{figure}
    \centering
\includegraphics[width=\linewidth]{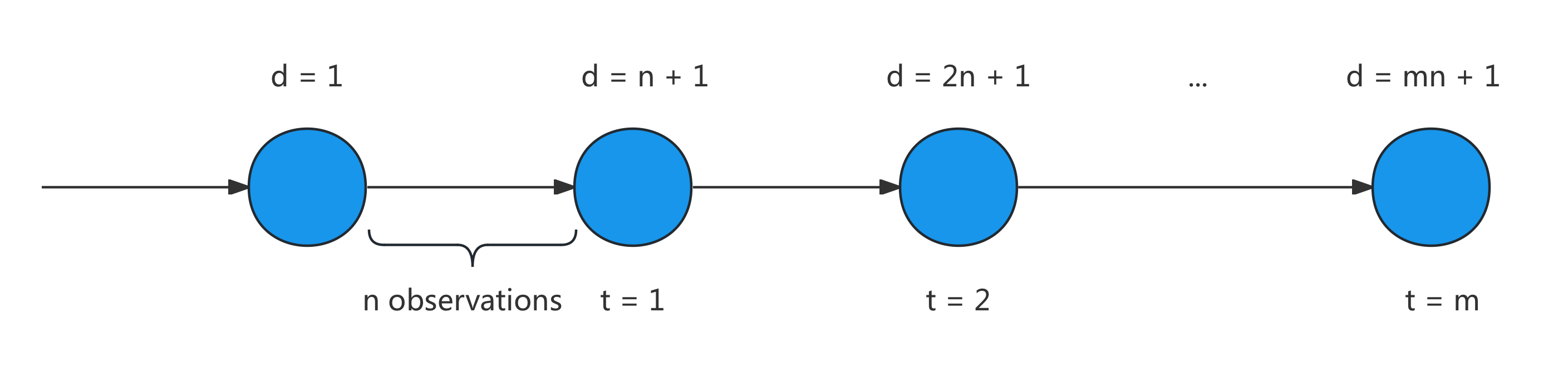}    \caption{First portfolio construction and reallocation diagram for multi-periods portfolios selection problems}
    \label{fig:enter-label}
\end{figure}

We consider an investor who aims to allocate wealth across $p$ risky assets over 
$m$ periods, assuming that asset returns are stationary within each period $t = 1,\ldots,m$. We denote the expected asset returns and the corresponding population covariance matrix in the $t$-th period by $\bmu_t$ and $\bSigma_t$, respectively.
Suppose an investor enters the market at day 1 with an initial wealth, allocates the wealth (constructing the first portfolio) at day $n+1$, where $n \in \mathcal{N}^+$ is a fixed number of trading days between two consecutive rebalances, and reallocates the wealth \footnote{For simplicity but without loss of generality, we assume the investor rebalances the portfolio at same frequency, e.g., annually as in Section \ref{sec:realdata}.} at the beginning of each of the following $m-1$ consecutive periods, where $m$ is the total number of portfolio formation decisions. At every decision day $d$, $d= n+1,2n+1,\ldots,mn+1$, we use the historical returns  $[\boldsymbol{R}_{d-n}, \boldsymbol{R}_{d-n+1},\ldots,\boldsymbol{R}_{d-1}]^{\top}\in \mathbb{R}^{n \times p}$, that contains observed returns of asset $j$ on its $j$-th column to estimate assets return and risk. Figure \ref{fig:enter-label} illustrates the decision process. 

Define the portfolio weight vector at each transaction time point, $t=1, \cdots, m$ as $\bw_t$
$$
\bw_t=\left(w_{t,1}, w_{t,2}, \ldots, w_{t,p}\right)^{\top},
$$
and the weights are taken at the beginning of period $t$. In real-world applications, investors tend to allocate wealth across a limited number of assets due to transaction costs, regulatory constraints, or interpretability concerns. To incorporate this preference, 
we consider the transaction cost under the Markowitz mean-variance framework \eqref{port_opt_exp}. And the optimal cost-aware portfolio at first portfolio construction stage ($t = 1$) is defined as 
\begin{equation}
\label{optimization}
\bw^{*}_1 := \mathop{\arg\min}_{\bw \in \mathbb{R}^{p}} \bw^{\top} \bSigma_1 \bw - \gamma\bw^{\top}\bmu_1 + \mathbf{C}_{1} (\bw) \quad  \text{s.t.} \ \bw^{\top} \textbf{1}=1,\ \|\bw\|_0 \leq s_0,
\end{equation}
where $\mathbf{C}_1 (\bw)$ is the transaction cost at first portfolio construction stage stage ($t = 1$), depending on the trading activity (changes in portfolio weights) and $s_0 < \infty$ denotes the maximum number of assets allowed in the portfolio. Similarly, the optimal cost-aware portfolio at reallocation stage ($t\ge 2$) is defined as
\begin{equation}
\label{reallocation optimization}
\bdelta^{*}_t := \mathop{\arg\min}_{\bdelta \in \mathbb{R}^{p}} \bdelta^{\top} \bSigma_t \bdelta + 2\bw^{+\top}_{t-1} \bSigma_t \bdelta_{t} - \gamma\bdelta^{\top}\bmu_t + \mathbf{C}_{t} (\bdelta) \quad  \text{s.t.} \ \bdelta^{\top} \textbf{1}=0,\ \|\bdelta\|_0 \leq s_0,
\end{equation}
 where the weight difference vector $\bdelta_t = \bw_t - \bw^+_{t-1}$, $\mathbf{C}_t (\bw)$ is the transaction cost at reallocation stage ($t\ge 2$), and $\bw^+_{t-1}$ means the pre-rebalancing weight as the value of portfolio changes over the holding period, and $\bw^+_{t-1} = (f_n\circ f_{n-1} \circ \dots \circ f_1)(\bw_{t-1})$, where $f_i(\bw) := \frac{\bw\odot(\mathbf{1} + \bR_{(t-2)n+i})}{1 + \bw^{\top}\bR_{(t-2)n+i}}$. This method is thus called the \emph{cost-aware portfolio estimator} (CAPE) in high dimensions. This paper proposes a new estimator targeting $\bw^{*}_1$ and $\bdelta^{*}_t$ under the high-dimensional model setting.

Denote the support sets of $\bw_1^{*}$ and $\bdelta_t^{*}$ as $\mathcal{A}_{1}=\left\{j: w^*_{1,j}\neq 0 \right\}$ and $\mathcal{A}_{t}=\left\{j: \delta^*_{t,j}\neq 0 \right\},\ t\geq 2$,  
 respectively. And we assume the cardinality $s_t=|\mathcal{A}_t| \leq s_0$, for $t = 1,2,\ldots,m$. We believe this assumption is empirically grounded as many active strategy investors only choose a relatively small number (compared to the vast universe) of assets considering the transaction and management fees.

To address the computational intractability of the $\ell_0$-constrained problems in
\eqref{optimization} and \eqref{reallocation optimization}, we approximate the $\ell_0$ constraint 
with a general sparsity-inducing penalty $p_\lambda(\cdot)$, yielding the following penalized formulations.

At the initial portfolio construction stage ($t=1$):
\begin{equation}
\label{optimization_penalty}
\hat{\bw}_1^{\mathrm{pen}} =
\mathop{\arg\min}_{\bw \in \mathbb{R}^{p}} 
\bw^{\top} \hat{\bSigma}_1 \bw 
- \gamma\, \bw^{\top} \hat{\bmu}_1 
+ \mathbf{C}_{1}(\bw) 
+ \sum_{j=1}^p p_\lambda(|w_j|)
\quad \text{s.t.} \quad \bw^{\top} \mathbf{1} = 1.
\end{equation}

At the reallocation stage ($t \ge 2$):
\begin{equation}
\label{optimization_penalty_reallocation}
\hat{\bdelta}_t^{\mathrm{pen}} =
\mathop{\arg\min}_{\bdelta \in \mathbb{R}^{p}} 
\bdelta^{\top} \hat{\bSigma}_t \bdelta 
- \gamma\, \bdelta^{\top} \hat{\bmu}_t 
+ \mathbf{C}_{t}(\bdelta) 
+ \sum_{j=1}^p p_\lambda(|\delta_j|)
\quad \text{s.t.} \quad \bdelta^{\top} \mathbf{1} = 0.
\end{equation}

Here, $p_\lambda(\cdot)$ denotes the sparsity-inducing penalty, and 
$\hat{\bmu}_t$ and $\hat{\bSigma}_t$ are the estimators of the mean vector $\bmu_t$ 
and covariance matrix $\bSigma_t$, respectively, for $t=1,2,\dots,m$.

Two popular choices of $p_\lambda(\cdot)$ are the $\ell_1$ penalty (i.e., Lasso), and the Smoothly Clipped Absolute Deviation (SCAD) penalty. While the Lasso penalty is convex and yields efficient computation, it introduces bias and requires the strong Irrepresentable Condition to achieve variable selection consistency \citep{zhao2006model}. In contrast, nonconvex penalties like SCAD can reduce the estimation bias and achieve better selection consistency under weaker conditions.

The SCAD penalty \citep{fan2001scadli} satisfies the following regularity conditions:
\begin{itemize}
\item[(i)] $P_{\lambda}(t)$ is increasing and concave on $t \in [0, \infty)$ with $P_{\lambda}(0) = 0$;
\item[(ii)] $P_{\lambda}$ is differentiable on $t \in (0, \infty)$ with $P_{\lambda}'(0+)\geq a_1 \lambda$;
\item[(iii)] $P_{\lambda}'(t) \geq a_1 \lambda$ for $t \in (0, a_2 \lambda]$;
\item[(iv)] $P_{\lambda}'(t)=0$ for $t \in [a \lambda, \infty)$ with some constant $a > a_2$,
\end{itemize}
where $a_1$ and $a_2$ are fixed positive constants.

The SCAD penalty has $a_1 = a_2 = 1$ and its derivative is given by
\begin{equation*}
P'_\lambda(|\tau|) = \lambda \left\{ I(|\tau| \leq \lambda) + \frac{(a \lambda - |\tau|)_+}{(a - 1)\lambda} I(|\tau| > \lambda) \right\}
\end{equation*}
for some $a > 2$, and $P'_\lambda(0+) = \lambda$.

Building on the general penalized formulations in \eqref{optimization_penalty} and \eqref{optimization_penalty_reallocation}, and leveraging the discussion of commonly used penalties such as Lasso and SCAD, we develop a novel cost-aware optimization framework specifically designed for high-dimensional portfolio selection. Specifically, we define the \emph{Cost-Aware Portfolio Estimator under SCAD Penalty} (CAPE-S) at the first portfolio construction stage as follows:
\begin{equation}
\begin{aligned}
\label{optimization_sample}
\hat{\bw}_1 :=
\mathop{\arg\min}_{\bw_1 \in \mathbb{R}^{p}} \; \bw_1^{\top} \hat{\bSigma}_1 \bw_1 - \gamma \bw_1^{\top} \hat{\bmu}_1 + \mathbf{C}_{1}(\bw_1) + P_{\lambda}(\bw_1) \quad \text{s.t.} \quad \bw_1^{\top} \mathbf{1} = 1,
\end{aligned}
\end{equation}
where \( P_{\lambda}(\bw_1) = \sum_{j=1}^p P_{\lambda}(\bw_{t,j}) \) is SCAD penalty. This formulation simultaneously accounts for portfolio risk, expected return, transaction cost, and sparsity, making it particularly suitable for high-dimensional financial settings.

To facilitate theoretical analysis, we consider the oracle estimator which assumes knowledge of the true support set \(\mathcal{A}_1\). The oracle portfolio weight estimator is defined by restricting the weights outside \(\mathcal{A}_1\) to zero:
\[
\hat{\bw}_1^{\mathrm{oracle}}  = \mathop{\arg\min}_{\bw_{\mathcal{A}_1^c} = \mathbf{0},\, \bw^{\top}_1 \mathbf{1} = 1} \bw^{\top}_1 \hat{\bSigma}_1 \bw_1 - \gamma \bw^{\top}_1 \hat{\bmu}_1 + \mathbf{C}_1(\bw_1).
\]

We assume the oracle solution satisfies the first-order optimality conditions:
\begin{equation*}
\label{oracle_first_order}
\nabla_j \mathcal{L}_{n,1}(\hat{\bw}_1^{\mathrm{oracle}}) + h^{\mathrm{oracle}} = 0, \quad \forall j \in \mathcal{A}_1,
\end{equation*}
where the Lagrange multiplier
\begin{align*}
    &h^{\mathrm{oracle}} = \frac{2 - \gamma \mathbf{1}^\top \hat\bSigma_{\mathcal{A}_1, \mathcal{A}_1}^{-1} \hat\bmu_{\mathcal{A}_1} + \mathbf{1}^\top \hat\bSigma_{\mathcal{A}_1, \mathcal{A}_1}^{-1}\mathbf{g}_{\mathcal{A}_1}}{\mathbf{1}^\top \hat\bSigma_{\mathcal{A}_1, \mathcal{A}_1}^{-1} \mathbf{1}},\\ 
    &\mathbf{g} :=\partial\mathbf{C}_1(\bw_1) / \partial \bw_1 \ \text{is the vector of sub-derivatives of} \ \mathbf{C}_1(\bw_1)
\end{align*}
and
\[
\mathcal{L}_{n,1}(\bw) = \bw^{\top} \hat{\bSigma}_1 \bw - \gamma \bw^{\top} \hat{\bmu}_1 + \mathbf{C}_1(\bw).
\]

The above describes the oracle estimator during the first initial portfolio construction stage. 
Similarly, for the reallocation stage with \( t \geq 2 \), the \emph{Cost-Aware Portfolio Estimator under SCAD Penalty (CAPE-S)} are defined as
\begin{equation*}
\begin{aligned}
\label{optimization_reallocation}
\hat{\bdelta}_t :=
\mathop{\arg\min}_{\bdelta_t \in \mathbb{R}^p} \; \bdelta^{\top}_t \hat{\bSigma}_t \bdelta_t + 2\bw^{+\top}_{t-1} \hat\bSigma_t \bdelta_{t} - \gamma \bdelta^{\top} \hat{\bmu}_t + \mathbf{C}_t(\bdelta_t) + P_{\lambda}(\bdelta_t) \quad \text{s.t.} \quad \bdelta^{\top}_t \mathbf{1} = 0,
\end{aligned}
\end{equation*}
where \( P_{\lambda}(\bdelta_t) = \sum_{j=1}^p P_{\lambda}(\bdelta_{t,j}) \) is SCAD penalty and the weight difference vector $\bdelta_t = \bw_t - \bw^+_{t-1}$, for $t\geq 2$, and \(\hat{\bmu}_t\) and \(\hat{\bSigma}_t\) denote estimators of the true mean vector \(\bmu_t\) and covariance matrix \(\bSigma_t\), respectively.

The corresponding oracle estimator is
\[
\hat{\bdelta}_t^{\mathrm{oracle}} = \mathop{\arg\min}_{\bdelta_{\mathcal{A}_t^c} = \mathbf{0},\, \bdelta^{\top}_t \mathbf{1} = 0} \bdelta^{\top}_t \hat{\bSigma}_t \bdelta_t + 2\bw^{+\top}_{t-1} \hat\bSigma_t \bdelta_{t} - \gamma \bdelta^{\top} \hat{\bmu}_t + \mathbf{C}_t(\bdelta_t),
\]
which satisfies
\begin{equation*}
\label{oracle_first_order_reallocation}
\nabla_j \mathcal{L}_{n,t}(\hat{\bdelta}_t^{\mathrm{oracle}}) + h_t^{\mathrm{oracle}} = 0, \quad \forall j \in \mathcal{A}_t,
\end{equation*}
where
\begin{align*}
    &h_t^{\mathrm{oracle}} = \frac{2 - \gamma \mathbf{1}^\top \hat\bSigma_{\mathcal{A}_t, \mathcal{A}_t}^{-1} \hat\bmu_{\mathcal{A}_t} + \mathbf{1}^\top \hat\bSigma_{\mathcal{A}_1, \mathcal{A}_1}^{-1}\mathbf{g}_{\mathcal{A}_t}}{\mathbf{1}^\top \hat\bSigma_{\mathcal{A}_t, \mathcal{A}_t}^{-1} \mathbf{1}},\\
    &\mathbf{g} :=\partial\mathbf{C}_t(\bdelta_t) / \partial \bdelta_t \ \text{is the vector of sub-derivatives of} \ \mathbf{C}_t(\bdelta_t),
\end{align*}
and
\[
\mathcal{L}_{n,t}(\bdelta_t) = \bdelta^{\top}_t \hat{\bSigma}_t \bdelta_t + 2\bw^{+\top}_{t-1} \hat\bSigma_t \bdelta_{t} - \gamma \bdelta^{\top}_t \hat{\bmu}_t + \mathbf{C}_t(\bdelta_t).
\]

However, the nonconvexity of SCAD poses significant computational challenges due to the presence of multiple local minima. To overcome this issue, \citet{zou2008one} proposed the local linear approximation (LLA) algorithm, which iteratively approximates the nonconvex problem by a sequence of weighted \(\ell_1\)-penalized problems, thereby inheriting the computational advantages of convex optimization. Rather than pursuing full convergence, they advocated for a properly initialized one-step LLA estimator, which is computationally efficient and shown to possess the strong oracle property under the condition that \(\bX^\top\bX/n \to \mathbf{M}\) for some positive definite matrix \(\mathbf{M}\). This framework was later extended by \citet{fan2014strong} to high-dimensional settings, demonstrating that initializing the LLA algorithm with a Lasso solution yields an oracle estimator with high probability. The algorithmic procedures for the first portfolio construction and subsequent reallocation stages are presented in Algorithm~\ref{lla-algorithm 1} and Algorithm~\ref{lla-algorithm 2}, respectively. Here, \(\hat{\bw}_1^{\text{initial}}\) and \(\hat{\bdelta}_t^{\text{initial}}\) denote initial estimators of \(\bw^*_1\) and \(\bdelta^*_t\), respectively, satisfying certain regularity conditions. The specific definitions of these initial estimators, tailored to different transaction cost structures, are provided in Sections~\ref{Quadratic Transaction Cost} and~\ref{Proportional Transaction Cost}.

\begin{algorithm}[htbp]
\caption{\small{Local Linear Approximation (LLA) Algorithm for First Portfolio Construction}}\label{lla-algorithm 1}
\begin{algorithmic}[1]
\State Initialize $\hat{\bw}_1^{(0)} = \hat{\bw}_1^\text{initial}$ and compute the adaptive weight:
$$
\hat{\btheta}^{(0)}=\left(\hat{\theta}_{1}^{(0)}, \ldots, \hat{\theta}_{p}^{(0)}\right)^{\top}=\left(P_{\lambda}^{\prime}\left(\left|\hat{w}_{1,1}^{(0)}\right|\right), \ldots, P_{\lambda}^{\prime}\left(\left|\hat{w}_{1,p}^{(0)}\right|\right)\right)^{\top}
$$
\For{$l=1,2,...$} \text{repeat the LLA iteration till convergence}
\State Obtain $\hat{\bw}_{t}^{(l)}$ by solving the following optimization problem 
    $$
    \hat{\bw}_1^{(l)}=\min _{\bw \in \mathbf{R}^{p},\bw^{\top}\mathbf{1}=1} \mathcal{L}_{n,1}(\boldsymbol{\bw})+\sum_{j} \hat{\theta}_{j}^{(l-1)} \cdot\left|w_{j}\right|
    $$
\State Update the adaptive weight vector $\hat{\boldsymbol{\btheta}}^{(l)}$ with $\hat{\theta}_j^{(l)} = P_{\lambda}^{\prime}\left(\left|\hat{w}_{1,j}^{(l)}\right|\right)$
\EndFor
\State \textbf{Output:} The portfolio allocation $\hat{\bw}_t^{\text{LLA}}$
\end{algorithmic}
\end{algorithm}

\begin{algorithm}[htbp]
\caption{\small{Local Linear Approximation (LLA) Algorithm for Reallocation}}\label{lla-algorithm 2}
\begin{algorithmic}[1]
\State Initialize $\hat{\bdelta}_t^{(0)} = \hat{\bdelta}_t^\text{initial}$ and compute the adaptive weight:
$$
\hat{\btheta}^{(0)}=\left(\hat{\theta}_{1}^{(0)}, \ldots, \hat{\theta}_{p}^{(0)}\right)^{\top}=\left(P_{\lambda}^{\prime}\left(\left|\hat{\delta}_{1,1}^{(0)}\right|\right), \ldots, P_{\lambda}^{\prime}\left(\left|\hat{\delta}_{1,p}^{(0)}\right|\right)\right)^{\top}
$$
\For{$l=1,2,...$} \text{repeat the LLA iteration till convergence}
\State Obtain $\hat{\bdelta}_{t}^{(l)}$ by solving the following optimization problem 
    $$
    \hat{\bdelta}_t^{(l)}=\min _{\bdelta \in \mathbf{R}^{p},\bdelta^{\top}\mathbf{1}=0} \mathcal{L}_{n,t}(\boldsymbol{\bdelta})+\sum_{j} \hat{\theta}_{j}^{(l-1)} \cdot\left|\delta_{j}\right|
    $$
\State Update the adaptive weight vector $\hat{\boldsymbol{\btheta}}^{(l)}$ with $\hat{\theta}_j^{(l)} = P_{\lambda}^{\prime}\left(\left|\hat{\delta}_{1,j}^{(l)}\right|\right)$
\EndFor
\State \textbf{Output:} The portfolio allocation $\hat{\bdelta}_t^{\text{LLA}}$
\end{algorithmic}
\end{algorithm}

To establish the non-asymptotic error bounds, we impose the following assumptions:
\begin{itemize}
\item[(A1)] There exist constants $C_1, C_2 > 0$, for $t = 1,2,\ldots,m$, such that
\begin{equation*}
\begin{aligned}
&\|\bSigma^{-1}_{\mathcal{A}_t,\mathcal{A}_t} \br\|_{\infty} < C_1, \quad \forall \br \in \{-1, 0, 1\}^{s_t}, \\
&|||\bSigma_{\mathcal{A}^c_t,\mathcal{A}_t}|||_\infty < C_2, \\
&0 < \lambda_{\min}(\bSigma_{\mathcal{A}_t,\mathcal{A}_t}) \leq \lambda_{\max}(\bSigma_{\mathcal{A}_t,\mathcal{A}_t}) < \infty.
\end{aligned}
\end{equation*}

\item[(A2)] For $t = 1,2,\ldots,m$ and some absolute constants $c_1, c_2, M_1, M_2$,
\begin{equation}\label{mean_var_max_bounds}
\begin{aligned}
\operatorname{Pr}\left(\|\hat{\bmu}_t - \boldsymbol{\mu}_t\|_{\infty} \leq M_1 \sqrt{\log p / n} \right) &\geq 1 - O(p^{-c_1}), \\
\operatorname{Pr}\left(\|\widehat{\boldsymbol{\Sigma}}_t - \boldsymbol{\Sigma}_t\|_{\max} \leq M_2 \sqrt{\log p / n} \right) &\geq 1 - O(p^{-c_2}).
\end{aligned}
\end{equation}

\item[(A3)] The estimated covariance matrix $\hat\bSigma_t$ satisfies the restricted strong convexity (RSC) condition:
\begin{equation}\label{RSC condition}
\forall \boldsymbol{\Delta} \in \mathbb{R}^p, \quad \boldsymbol{\Delta}^{\top} \hat{\bSigma}_t \boldsymbol{\Delta} \geq v \|\boldsymbol{\Delta}\|_2^2 - \tau \sqrt{\frac{\log p}{n}} \|\boldsymbol{\Delta}\|_1,
\end{equation}
where $v > 0$, $\tau \geq 0$, and $t = 1,2,\ldots,m$.
\end{itemize}

\begin{remark} 
\textbf Assumption (A1) guarantees three key properties. The bound on $\|\bSigma^{-1}_{\mathcal{A}_t,\mathcal{A}_t} \br\|_\infty$ ensures that the $\ell_1$-norms of $\bw_1^*$ and $\bdelta_t^*$ are controlled, preventing divergence in the estimation process. This condition aligns with assumptions in \citet{cai2024asset}. The constraint on $|||\bSigma_{\mathcal{A}^c_t,\mathcal{A}_t}|||_\infty$, while weaker than the irrepresentable condition required for the Lasso, limits the correlation between active and inactive variables, promoting both sparsity and stability. Finally, the eigenvalue condition guarantees that $\bSigma_{\mathcal{A}_t,\mathcal{A}_t}$ is well-conditioned, which facilitates stable matrix inversion and avoids ill-posedness when estimating coefficients restricted to the active set $\mathcal{A}_t$.

Assumption (A2) holds when the sample mean vector and sample covariance matrix are used as estimators, provided that the returns $\boldsymbol{R}_{(t-1)n+1}, \ldots, \boldsymbol{R}_{tn}$ are i.i.d. sub-Gaussian vectors with mean $\boldsymbol{\mu}_t$ and covariance matrix $\boldsymbol{\Sigma}_t$, and that $\log p < n$, as shown in \citet{ProjectionTest}. Furthermore, we show that the linear shrinkage estimator \citep{ledoit2020analytical} for the population covariance matrix also satisfies (A2) under the same sub-Gaussianity assumption and mild structural conditions on $\boldsymbol{\Sigma}_t$ in the supplementary material. The linear shrinkage estimator has more favorable properties and, in high-dimensional settings (e.g., $p$ comparable to or exceeding $n$), provides well-conditioned covariance estimates that make the classical Markowitz mean–variance model \eqref{baseline1} and its transaction-cost extension \eqref{baseline3} feasible and numerically stable. Moreover, the bounds in \eqref{mean_var_max_bounds} can be extended to settings with dependent observations, including strong-mixing ($\alpha$-mixing) processes, $m$-dependence, and time series exhibiting physical (or functional) dependence \citep{Chang_2024}. For simplicity, we focus on the i.i.d.\ case in this paper and defer the dependent setting to future work.

Assumption (A3) is a standard requirement in high-dimensional estimation. \citet{RegularizedMestimators} established that the sample covariance matrix satisfies the RSC condition with high probability under sub-Gaussian assumptions. We further show that both linear and nonlinear shrinkage estimators \citep{ledoit2020analytical} satisfy (A3), as detailed in the supplementary material.
\end{remark}

\subsection{CAPE-S with Quadratic Transaction Cost} \label{Quadratic Transaction Cost}


\subsubsection{First Portfolio Construction}

We begin with considering the quadratic transaction cost from the initial period, $t=1$,
$$\mathbf{C}_1 (\bw) = (\bm{\beta}\odot \bw)^{\top}\bw$$
where $\bm{\beta}=(\beta_1,...,\beta_p)^{\top}, \beta_j> 0$ is the cost parameter, which means the unit transaction cost of the risky asset $j$ is a linear function of the trading size with slope $\beta_j$, for $j=1,...,p$. 
The problem \eqref{optimization} then becomes
\begin{equation}\label{min w}
    \begin{aligned}
    \min_{\bw_1} \bw_1^{\top} \bSigma_1 \bw_1  - \gamma\bw_1^{\top}\bmu_1 + (\bm{\beta}\odot \bw_1)^{\top}\bw_1,\quad \text{ s.t. } \bw_1^{\top}\mathbf{1} = 1,\ \|\bw_1\|_0\le s_0.
    \end{aligned}
\end{equation}

This is a constrained convex optimization problem. By applying the method of Lagrange multipliers and using the definitions of \(\bw^*_1\) and the active set \(\mathcal{A}_1\), we obtain the explicit solution
\[
\bw_{\mathcal{A}_1}^{*}   =  \frac{1}{2} \tilde{\bSigma}_{\mathcal{A}_1,\mathcal{A}_1}^{-1}\left(\gamma \bmu_{\mathcal{A}_1} + \frac{2 - \gamma \mathbf{1}^{\top} \tilde{\bSigma}_{\mathcal{A}_1,\mathcal{A}_1}^{-1} \bmu_{\mathcal{A}_1}}{\mathbf{1}^{\top} \tilde{\bSigma}_{\mathcal{A}_1,\mathcal{A}_1}^{-1} \mathbf{1}} \mathbf{1}\right) \ \text{and} \ \bw_{\mathcal{A}^{c}_1}^{*} = \mathbf{0}.
\]
Here, \(\tilde{\bSigma}_1 = \bSigma_1 + \mathbb{B}\), with \(\mathbb{B} = \mathrm{diag}(\beta_1, \ldots, \beta_p)\). The explicit solution above is derived using the method of Lagrange multipliers and the sparsity structure of the optimizer. For a detailed derivation, we refer the reader to the proof of Theorem~\ref{thm1} in the supplementary material.

As discussed earlier, we employ the LLA algorithm to solve CAPE-S and use the CAPE-L $\widehat{\boldsymbol{\bw}}_1^{\text{Lasso},\beta}$ as the initial estimate required by the LLA algorithm. The following lemma establishes the consistency of the CAPE-L.

\begin{lemma}
\label{Lasso_estimation_error}
For the first portfolio construction period, suppose that Assumptions (A1)-(A3) hold and  $\|\boldsymbol{\beta}\|_{\infty} \ll 1/C_1$, where $C_1$ is defined in Assumption (A1). Let $\widehat{\boldsymbol{\bw}}_1^{\text{Lasso},\beta}$ be the unique optimum of the program
\begin{equation}
\label{Lasso_initial_qua}
\widehat{\boldsymbol{\bw}}_1^{\text{Lasso},\beta} = \mathop{\arg\min}_{\mathbf{1}^{\top}\bw_1=1} \bw_1^{\top} \hat\bSigma_1 \bw_1  - \gamma\bw_1^{\top}\hat\bmu_1 + (\bm{\beta}\odot \bw_1)^{\top}\bw_1 + \lambda_{\text{Lasso},1}\|\bw_1\|_{1}
\end{equation}
 with $\lambda_{\text{Lasso},1} = M \sqrt{\frac{\log p}{n}}$
for some large constant $M$. Then with probability at least $1-c_1 p^{-c_2}$ for some constant $c_1,c_2$, we have 
$$\|\widehat{\boldsymbol{\bw}}_1^{\text{Lasso},\beta}-\bw_{1}^{*}\|_{1}=O\left(s_1\sqrt{\frac{\log p}{n}}\right) \quad \text{and}\quad \|\widehat{\boldsymbol{\bw}}_1^{\text{Lasso},\beta}-\bw_1^{*}\|_{2}=O\left(\sqrt{\frac{s_1 \log p}{n}}\right).$$

\end{lemma}

In Lemma \ref{Lasso_estimation_error}, the estimator $\widehat{\boldsymbol{\bw}}_1^{\text{Lasso},\beta}$ defined in Equation \eqref{Lasso_initial_qua} essentially applies the $\ell_1$ penalty to the penalty term introduced in Equation \eqref{optimization_penalty}. Following the naming convention used in Equation \eqref{optimization_sample}, this class of estimators can be referred to as the \emph{Cost-Aware Portfolio Estimator under Lasso Penalty} (CAPE-L) at the first portfolio construction stage. Lemma \ref{Lasso_estimation_error} says that the initial Lasso estimator yields an estimate that is consistent in $\ell_2$ norm if $\sqrt{s_1 \log p/n} \to 0$ as $n,p \to \infty$. In other words, if the portfolio has sufficient sparsity that $s_1 = o(\sqrt{\frac{n}{\log p}})$, the estimation error can be ignored. Our error rate also matches the rate of Lasso estimator for linear regression. 

For the first portfolio construction, the oracle portfolio construction weight estimator with quadratic transaction cost is given as 
\begin{equation}\label{oracle w beta}
\begin{aligned}
\hat{\boldsymbol{\bw}}_{1}^{\text {oracle},\beta}
 = & \mathop{\arg\min}_{ \boldsymbol{\bw}_{\mathcal{A}_{1}^{c}}=\mathbf{0},\bw^{\top}_1\mathbf{1}=1} \bw^{\top}_{1}\widehat{\bSigma}_{1}\bw_{1} - \gamma\bw^{\top}_{1}\hat\bmu_{1} + (\boldsymbol{\beta}\odot\bw_{1})^{\top}\bw_{1}.
\end{aligned}
\end{equation}

The following theorem states that the oracle estimator can be obtained after two iterations by implementing the LLA algorithm to solve CAPE-S, and it also provides the convergence rate in the $\ell_{\infty}$-norm.

\begin{theorem}
\label{thm1}
Suppose that Assumptions (A1)–(A3) hold and that $\|\boldsymbol{\beta}\|_{\infty} \ll 1/C_1$, where $C_1$ is defined in Assumption (A1). Choose $\lambda \geq M\sqrt{\frac{s_1 \log p}{n}}$ for some sufficiently large constant $M$. Let $\widehat{\bw}^{\mathrm{LLA},\beta}_1$ denote the solution obtained from the LLA algorithm~\ref{lla-algorithm 1} with the SACD penalty, initialized at $\hat{\boldsymbol{\bw}}_{1}^{\mathrm{Lasso},\beta}$. If $s_1 = O\!\left(\sqrt{\frac{n}{\log p}}\right)$ and $\|\bw^{*}_1\|_{\min} > (a+1) \lambda_1$, then $\widehat{\bw}^{\mathrm{LLA},\beta}_1$ converges to $\hat{\boldsymbol{\bw}}_1^{\mathrm{oracle},\beta}$ within two iterations with probability at least $1 - c_1 k^{-c_2}$ for all $k \in [s_1, p]$ and some constants $c_1, c_2 > 0$. Furthermore,
\[
\|\widehat{\boldsymbol{\bw}}_1^{\mathrm{LLA},\beta} - \bw_{1}^{*}\|_{\infty} = O\!\left(\sqrt{\frac{\log k}{n}}\right).
\]
\end{theorem}

Theorem \ref{thm1} establishes that, under mild regularity conditions, the proposed CAPE-S estimator attains the same performance as the oracle estimator $\widehat{\bw}_1^{\mathrm{oracle},\beta}$ after only two iterations of the LLA algorithm. In particular, when the portfolio dimension $k$ diverges, the two estimators coincide with high probability, ensuring that CAPE-S inherits the oracle’s optimality properties in large-scale settings. This oracle property provides the theoretical foundation for subsequent results on risk and return estimation.

\begin{corollary}
\label{sp_corollary}
Given the same conditions as in Theorem \ref{thm1}, if $s_1 \sqrt{ \log s_1 / n} = o(1) $, then we have
$$
|(\hat{\bw}^{\text{LLA},\beta}_1)^{\top}\widehat{\bmu}_1-(\bw_1^{*})^{\top} \bmu_1 |=o_{p}(1), \qquad |(\hat{\bw}^{\text{LLA},\beta}_1)^{\top}\widehat{\bmu}_2-(\bw_1^{*})^{\top} \bmu_2 |=o_{p}(1) ;
$$
$$| (\widehat{{\bw}}_{1}^{\text{LLA},\beta})^{\top} \widehat{\bSigma}_1\hat{\bw}^{\text{LLA},\beta}_1-\bw_{1}^{*} \bSigma_1 \bw_1^{*} |=o_{p}(1), \qquad 
| (\widehat{{\bw}}_{1}^{\text{LLA},\beta})^{\top} \widehat{\bSigma}_2\hat{\bw}^{\text{LLA},\beta}_1-(\bw_{1}^{*})^{\top} \bSigma_2 \bw_1^{*} |=o_{p}(1);
$$
$$ \left|\frac{(\hat{\bw}^{\text{LLA},\beta}_1)^{\top}\widehat{\bmu}_{1}}{\sqrt{(\hat{\bw}^{\text{LLA},\beta}_1)^{\top}\widehat{\bSigma}_1\widehat{\boldsymbol{\bw}}_1}} -\frac{(\bw_1^{*})^{\top} \bmu_1}{\sqrt{(\bw_{1}^{*})^{\top} \bSigma_1 \bw_1 }} \right|=o_{p}(1) ,\qquad 
\left|\frac{(\hat{\bw}^{\text{LLA},\beta}_1)^{\top}\widehat{\bmu}_{2}}{\sqrt{(\hat{\bw}^{\text{LLA},\beta}_1)^{\top} \widehat{\bSigma}_2\widehat{\boldsymbol{\bw}}_1}} -\frac{(\bw_1^{*})^{\top} \bmu_2}{\sqrt{(\bw_{1}^{*})^{\top} \bSigma_2 \bw_1 }} \right|=o_{p}(1) .
$$
\end{corollary}

Corollary \ref{sp_corollary} shows that if the sparsity level and the dimension satisfy that $s_1 \sqrt{\log s_1 / n} = o(1)$, the in-sample and out-of-sample Sharpe ratio estimation error of initial period vanishes when $n$ goes to infinity.  \cite{CANER2023393} established similar results under a factor model framework. In contrast, we impose a weaker condition on the data-generating process. 

\subsubsection{Portfolio Reallocation}
At reallocation decision day $d$, the transaction costs $C_t(\bw)$ are a quadratic function of the weight difference vector $\bdelta_t$, defined as $\bdelta_t:= \bw_{t} - \bw^+_{t-1}$, for $t\ge 2$. And $\bdelta_t$ represents the difference between the new allocation $\bw_{t}$ and the market value-adjusted allocation $\bw^+_{t-1}$. The form of transaction costs $C_t(\bw)$ is thus given by
$$\mathbf{C}_t (\bw) = [\bm{\beta}\odot (\bw_{t} -\bw^+_{t-1})]^{\top} (\bw_{t} -\bw^+_{t-1})$$
The program \eqref{reallocation optimization} turns out to be 

\begin{equation}\label{min delta}
\begin{aligned}
\min_{\bdelta_{t}}\ \bdelta^{\top}_{t} \bSigma_t \bdelta_{t} + 2\bw^{+\top}_{t-1} \bSigma_t \bdelta_{t}  -\gamma \bdelta_{t}^{\top}\bmu_t + (\bm{\beta}\odot \bdelta_{t})^{\top} \bdelta_{t}, \text{ s.t. } \bdelta_{t}^{\top}\mathbf{1} =  0,\ \| \bdelta_t \|\le s_0.
\end{aligned}
\end{equation}
Similarly, using the Lagrange multiplier method, we have that  
$$
\bdelta_{\mathcal{A}_t}^{*}  =  \frac{1}{2} 
\tilde{\bSigma}_{\mathcal{A}_t,\mathcal{A}_t}^{-1}\left(\gamma \bmu_{\mathcal{A}_t} - 2(\bSigma_t\bw^{+}_{t-1})_{\mathcal{A}_t} + \frac{2 - \gamma \mathbf{1}^{\top} \tilde{\bSigma}_{\mathcal{A}_t,\mathcal{A}_t}^{-1} \bmu_{\mathcal{A}_t}}{\mathbf{1}^{\top} \tilde{\bSigma}_{\mathcal{A}_t,\mathcal{A}_t}^{-1} \mathbf{1}} \mathbf{1}\right)\ \text{and} \ \bdelta_{\mathcal{A}^{c}_t}^{*} = \mathbf{0}. 
$$


In high-dimensional regimes, we consider the following constrained and
regularized quadratic programming to encourage a sparse estimator of $\bdelta_t^{*}$
\begin{equation*}
\begin{aligned}
\min_{\bdelta_{t}}\ \bdelta^{\top}_{t} 
\hat{\tilde{\bSigma}}_t \bdelta_{t}  + 2\bw_{t-1} ^{+\top}\hat{\bSigma}_t\bdelta_{t}  -\gamma \bdelta_{t}^{\top}\hat{\bmu}_t + P_{\lambda}(\bdelta_{t}),\text{ s.t. } \bdelta_{t}^{\top}\mathbf{1} =  0, 
\label{dynamic} 
\end{aligned}
\end{equation*}
where $\hat{\bSigma}_t$ and $\hat{\bmu}_t$ are the estimators of covariance matrix and mean of the asset returns of the $t$ period, $\hat{\tilde{\bSigma}}_t =  \hat{\bSigma}_t +\mathbb{B}$. By incorporating the transaction costs in the mean variance problem and regularizing rebalancing events, our portfolio achieves a higher post-cost Sharpe ratio and reduced transaction costs, as demonstrated in numerical studies.

Define the oracle estimator (with known true support of $\bdelta$) with quadratic transaction cost under portfolio reallocation as 
\begin{equation}\label{oracle delta beta}
\begin{aligned}
\hat{\boldsymbol{\bdelta}}_t^{\text {oracle},\beta}
= & \mathop{\arg\min}_{ \boldsymbol{\bdelta}_{\mathcal{A}_{t}^{c}}=\mathbf{0},\bdelta^{\top}\mathbf{1}=0} \bdelta^{\top}_{t}\widehat{\tilde{\bSigma}}_{t}\bdelta_{t} + 2\bw_{t-1}^{+\top}\widehat{\bSigma}_t \bdelta_{t} - \gamma\bdelta^{\top}_{t}\hat\bmu_{t}.
\end{aligned}
\end{equation}
where \(\hat{\tilde{\bSigma}}_t = \hat\bSigma_t + \mathbb{B}\), with \(\mathbb{B} = \mathrm{diag}(\beta_1, \ldots, \beta_p)\).
\begin{lemma}
\label{thm2 lemma}
    Suppose that Assumptions (A1)-(A3) hold and  $\|\boldsymbol{\beta}\|_{\infty} \ll 1/C_1$ with $C_1$ defined in Assumption (A1), let $\widehat{\boldsymbol{\bdelta}}_t^{\text{Lasso},\beta}$ be the unique optimum of the program  
    \begin{equation}
    \label{initial beta}
        \widehat{\boldsymbol{\bdelta}}_t^{\text{Lasso},\beta} = \mathop{\arg\min}_{\mathbf{1}^{\top}\bdelta_t=0} \bdelta^{\top}_{t} \hat{\tilde{\bSigma}}_t \bdelta_{t}  + 2\bw_{t-1} ^{+\top}\hat{\bSigma}_t\bdelta_{t}  -\gamma \bdelta_{t}^{\top}\hat{\bmu}_t + \lambda_{\text{Lasso},t}\|\bdelta_t\|_{1}.
    \end{equation}
\noindent with $\lambda_{\text{Lasso},t} = M \sqrt{\frac{\log p}{n}}$
for some large constant $M$. Then with probability at least $1-c_1 p^{-c_2}$ for some constant $c_1,c_2$, we have
    $$
    \|\widehat{\boldsymbol{\bdelta}}_t^{\text{Lasso},\beta}-\bdelta_{t}^{*}\|_{1}=O_p\left(s_t\sqrt{\frac{\log p}{n}}\right)\quad  \text{and}\quad \|\widehat{\boldsymbol{\bdelta}}_t^{\text{Lasso},\beta}-\bdelta_t^{*}\|_{2}=O_p\left(\sqrt{\frac{s_t\log p}{n}}\right).
    $$
\end{lemma}

\begin{theorem}
\label{thm2}
Suppose that Assumptions (A1)-(A3) hold, and that $\|\boldsymbol{\beta}\|_{\infty} \ll 1/C_1$, where $C_1$ is defined in Assumption (A1). Let $s_t = O\left(\sqrt{\frac{n}{\log p}}\right)$, and assume that $\left\|\bdelta^{*}_t\right\|_{\min} > (a+1) \lambda_t$. Choose $\lambda_t \geq M \sqrt{\frac{s_t \log p}{n}}$ for some sufficiently large constant $M$. Let $\widehat{\bdelta}^{\text{LLA},\beta}_t$ be the optimum of the LLA algorithm~\ref{lla-algorithm 2} with SCAD penalty,  initialized at $\hat{\boldsymbol{\bdelta}}_{t}^{\text {Lasso},\beta}$, as defined in \eqref{initial beta}. Then, we have the following results:

\begin{itemize}
    \item[(i)] If $\|\bw^+_{t-1}\|_{1} < C_0$, then $\widehat{\bdelta}^{\text{LLA},\beta}_t$ converges to $\hat{\boldsymbol{\bdelta}}_{t}^{\text {oracle},\beta}$ after two iterations with probability at least $1 - c_1 p^{-c_2}$ for some constants $c_1, c_2 > 0$. Furthermore, 
    \[
    \|\widehat{\boldsymbol{\bdelta}}_t^{\text{LLA},\beta} - \bdelta_{t}^{*}\|_{\infty} = O\left(\sqrt{\frac{\log p}{n}}\right).
    \]

    \item[(ii)] If $\|\bw^+_{t-1,\mathcal{Q}}\|_1 \leq C_0$ and $\bw^+_{t-1,\mathcal{Q}^c} = \mathbf{0}$ for some set $\mathcal{Q} \subset [1, \ldots, p]$ and absolute constant $C_0$, then $\widehat{\bdelta}^{\text{LLA},\beta}_t$ converges to $\hat{\boldsymbol{\bdelta}}_{t}^{\text {oracle},\beta}$ after two iterations with probability at least $1 - c_1 k^{-c_2}$ for all $k \in [\max\{s_t,|\mathcal{Q}|\}, p]$ and some constants $c_1, c_2 > 0$. Furthermore, 
    \[
    \|\widehat{\boldsymbol{\bdelta}}_t^{\text{LLA},\beta} - \bdelta_{t}^{*}\|_{\infty} = O\left(\sqrt{\frac{\log k}{n}}\right).
    \]
\end{itemize}
\end{theorem}

Theorem~\ref{thm2} and Lemma~\ref{thm2 lemma} have implied that in portfolio reallocation, the LLA algorithm initialized at CAPE-L yields  a local optimum of the nonconvex regularized problem that satisfies oracle properties with low $\ell_{\infty}$ error.  Theorem \ref{thm2} highlights the impact of different estimation methods employed in the previous $t-1$ stages on the performance of our proposed method in stage $t$. Specifically, if regularized estimation methods, such as the proposed CAPE-S in \eqref{optimization_sample} or CAPE-L in \eqref{Lasso_initial_qua}, were consistently applied in the previous 
$t-1$ stages, the resulting $\bw^+_{t-1}$ will exhibit a certain level of sparsity. This sparsity enables $\widehat{\bdelta}^{\text{LLA},\beta}_t$ to achieve a superior convergence rate in the current stage. Conversely, if the estimation methods used in the earlier stages fail to produce sparse estimates, leading to a non-sparse $\bw^+_{t-1}$, our method can still ensure that $\widehat{\bdelta}^{\text{LLA},\beta}_t$ converges to 
$\bdelta^*_t$ with probability approaching 1. However, the convergence rate will be $\sqrt{\log p/n}$, which is attributable to the lack of sparsity in $\bw^+_{t-1}$. Corollary 2 provides detailed conclusions regarding the estimation of $\bw^*_t$ when the proposed CAPE-S method is consistently applied across the previous $t-1$ stages.

\begin{corollary}
\label{sp_corollary_reallocate}
Define $s_0\leq k_t\leq p$ and $\widehat{\bw}_t^{\text{LLA},\beta} = \bw_{t-1}^+ + \widehat{\bdelta}_t^{\text{LLA},\beta}$, and we use CAPE-S to make estimation in the past $t-1$ stage. Given the same conditions as in Theorem \ref{thm1} and Theorem \ref{thm2}, with $k_t \sqrt{ \log k_t / n} = o(1) $, then we have
$$
\|\widehat{\boldsymbol{\bw}}_t^{\text{LLA},\beta}-\bw_{t}^{*}\|_{\infty}=O_p\left(\sqrt{\frac{\log k_t}{n}}\right);
$$
$$
|(\widehat{\boldsymbol{\bw}}^{\text{LLA},\beta}_t)^{\top}\widehat{\bmu}_t-(\bw_t^{*})^{\top} \bmu_t |=o_{p}(1), \qquad |(\widehat{\boldsymbol{\bw}}^{\text{LLA},\beta}_t)^{\top}\widehat{\bmu}_{t+1}-(\bw_t^{*})^{\top} \bmu_{t+1} |=o_{p}(1) ;
$$
$$| (\widehat{\bw}^{\text{LLA},\beta}_{t})^{\top} \widehat{\bSigma}_t\widehat{\boldsymbol{\bw}}^{\text{LLA},\beta}_t-(\bw_{t}^{*})^{\top} \bSigma_t \bw_t^{*} |=o_{p}(1), \qquad 
| (\widehat{{\bw}}^{\text{LLA},\beta}_{t})^{\top} \widehat{\bSigma}_{t+1}\widehat{\boldsymbol{\bw}}^{\text{LLA},\beta}_t-(\bw_{t}^{*})^{\top} \bSigma_{t+1} \bw_t^{*} |=o_{p}(1);
$$
$$ \left|\frac{(\widehat{\boldsymbol{\bw}}^{\text{LLA},\beta}_t)^{\top}\widehat{\bmu}_{t}}{\sqrt{(\widehat{\boldsymbol{\bw}}^{\text{LLA},\beta}_t)^{\top} \widehat{\bSigma}_t\widehat{\boldsymbol{\bw}}^{\text{LLA},\beta}_t}} -\frac{(\bw_t^{*})^{\top} \bmu_t}{\sqrt{(\bw_{t}^{*})^{\top} \bSigma_t \bw_t^{*} }} \right|=o_{p}(1), \  
\left|\frac{(\widehat{\boldsymbol{\bw}}^{\text{LLA},\beta}_t)^{\top}\widehat{\bmu}_{t+1}}{\sqrt{(\widehat{\boldsymbol{\bw}}^{\text{LLA},\beta}_t)^{\top} \widehat{\bSigma}_{t+1}\widehat{\boldsymbol{\bw}}^{\text{LLA},\beta}_t}} -\frac{\bw_t^{*} \bmu_{t+1}}{\sqrt{(\bw_{t}^{*})^{\top} \bSigma_{t+1} \bw^*_t }} \right|=o_{p}(1) .
$$
\end{corollary}
Corollary \ref{sp_corollary_reallocate} 
shows that if the sparsity level and the dimension satisfy that $k_t \sqrt{\log k_t / n} = o(1) $, the in-sample and out-of-sample Shape Ratio estimation errors of reallocation periods vanish when $n$ goes to infinity. 

\subsection{CAPE-S with Proportional Transaction Cost}
\label{Proportional Transaction Cost}


\subsubsection{First Portfolio Construction}

We then consider some real trading scenarios where costs are proportional to the sum of absolute rebalancing ($\ell_1$-norm of rebalancing). For the proportional transaction cost, the optimization problem corresponding to the initial period is as follows: 
\begin{equation*}
\begin{aligned}
\min_{\bw_1}\ \bw^{\top}_1 \bSigma_1 \bw_1 - \gamma\bw_1^{\top}\bmu_1 + \|\boldsymbol{\alpha} \odot \bw_1\|_1, \text{ s.t. } \bw_{1}^{\top}\mathbf{1} =  1, \ \|\bw_1\|_0\le s_0, \label{population_proportional}
\end{aligned}
\end{equation*}
where $\bm{\alpha}=(\alpha_1,...,\alpha_p)^{\top}, \alpha_j> 0$ is the cost parameter, which is similar to $\boldsymbol{\beta}$ in Section \ref{Quadratic Transaction Cost}.
By Lagrange multiplier method and the definition of $\bw^*_1$ and $\mathcal{A}_1$, solving for $\bw_1^{*}$ yields
$$
\bw_{\mathcal{A}_1}^{*}   =  \frac{1}{2} \bSigma^{-1}_{\mathcal{A}_1,\mathcal{A}_1}\left(\gamma \bmu_{\mathcal{A}_1} -\boldsymbol{\alpha}\odot\mathbf{g}_{\mathcal{A}_1}+ \frac{2 - \gamma \mathbf{1}^{\top} \bSigma^{-1}_{\mathcal{A}_1,\mathcal{A}_1} \bmu_{\mathcal{A}_1} + \mathbf{1}^{\top} \bSigma^{-1}_{\mathcal{A}_1,\mathcal{A}_1} (\boldsymbol{\alpha}\odot\mathbf{g}_{\mathcal{A}_1})}{\mathbf{1}^{\top} \bSigma^{-1}_{\mathcal{A}_1,\mathcal{A}_1} \mathbf{1}} \mathbf{1}\right)\quad \text{and}\quad \bw^{*}_{\mathcal{A}^c_1} = \mathbf{0}.
$$
where $\mathbf{g}$ is the vector of sub-derivatives of $\left\|\bw_1\right\|_{1}$, i.e., $\mathbf{g}:=\partial\left\|\bw_1\right\|_{1} / \partial \bw_1$, consisting of elements which are 1 or $-1$ in case $w_j>0$ or $w_j<0$, respectively, or $\in[-1,1]$ in case $w_j=0$.
Similar to Lemma \ref{Lasso_estimation_error} in Section \ref{Quadratic Transaction Cost}, Lemma \ref{Lasso_estimation_error_proportional}  shows the consistency of the CAPE-L $\widehat{\boldsymbol{\bw}}_1^{\text{Lasso},\alpha}$ under proportional transaction cost.
\begin{lemma}
\label{Lasso_estimation_error_proportional}
For the initial period, suppose that Assumptions (A1)-(A3) hold, let $\widehat{\boldsymbol{\bw}}_1^{\text{Lasso},\alpha}$ be the unique optimum of the program 
\begin{equation}
\label{Lasso_initial_pro}
\widehat{\boldsymbol{\bw}}_1^{\text{Lasso},\alpha} = \mathop{\arg\min}_{\mathbf{1}^{\top}\bw_1=1} \bw_1^{\top} \hat{\bSigma}_1 \bw_1 - \gamma \bw_1^{\top}\hat{\bmu}_1 +   \Vert \boldsymbol{\alpha}\odot\bw_1\Vert_1 + \lambda_{\text{Lasso},1}\|\bw_1\|_{1},
\end{equation}
 with $\lambda_{\text{Lasso},1} = M \sqrt{\frac{\log p}{n}}$ for some large constant $M$, then with probability at least $1-c_1 p^{-c_2}$ for some constants $c_1,c_2$, we have 
$$\|\widehat{\boldsymbol{\bw}}_1^{\text{Lasso},\alpha}-\bw_{1}^{*}\|_{1} = O\left(s_1\|\balpha\|_{\infty} + s_1\sqrt{\frac{\log p}{n}} \right)\quad \text{and}
 \quad \|\widehat{\boldsymbol{\bw}}_1^{\text{Lasso},\alpha}-\bw_1^{*}\|_{2} = O\left(\sqrt{s_1}\|\balpha\|_{\infty} + \sqrt{\frac{s_1\log p}{n}}\right).$$
\end{lemma}
Lemma \ref{Lasso_estimation_error_proportional} differs from Lemma \ref{Lasso_estimation_error} in Section \ref{Quadratic Transaction Cost} due to the distinct effects of the quadratic and proportional transaction costs. Specifically, the quadratic transaction cost can be interpreted as a perturbation to the diagonal elements of the covariance matrix. In contrast, the proportional transaction cost has a different influence, which is reflected in the convergence rate of $\widehat{\bw}^{\text{Lasso},\alpha}_1$. If the convergence rate of $\alpha$ exceeds that of $\lambda_{\text{Lasso}}$, the convergence rate of $\widehat{\bw}^{\text{Lasso},\alpha}_1$ will be determined by $\alpha$.

Similar to \eqref{oracle w beta}, at first portfolio construction stage, we define the oracle portfolio construction weight estimator with proportional transaction cost as 
\begin{equation}\label{oracle w alpha}
\begin{aligned}
\hat{\boldsymbol{\bw}}_{1}^{\text {oracle},\alpha}
 =  \mathop{\arg\min}_{ \boldsymbol{\bw}_{\mathcal{A}_{1}^{c}}=\mathbf{0},\bw^{\top}\mathbf{1}=1} \bw^{\top}_{1}\widehat{\bSigma}_{1}\bw_{1} - \gamma\bw^{\top}_{1}\hat\bmu_{1} + \|\boldsymbol{\alpha}\odot\bw_{1}\|_1.
\end{aligned}
\end{equation}

\begin{theorem}\label{thm3}
Under Assumptions (A1)-(A3) and $\|\boldsymbol{\alpha}\|_{\infty} \leq C\sqrt{\frac{\log p}{n}}$ for some large constant $C$, we select $\lambda \geq M\sqrt{\frac{s_1\log p}{n}}$ for some large constant $M$. Let $\widehat{\bw}^{\text{LLA},\alpha}_t$ be the optimum of LLA algorithm~\ref{lla-algorithm 1} with SCAD penalty, initialized at $\hat{\boldsymbol{\bw}}_{t}^{\text {Lasso },\alpha}$. If $s_1 = O\left(\sqrt\frac{n}{\log p}\right)$ and $\left\|\bw^{*}_1\right\|_{\min }>(a+1) \lambda_1$, $\widehat{\bw}^{\text{LLA},\alpha}_t$ converges to $\hat{\boldsymbol{\bw}}^{\text {oracle},\alpha}$ after two iterations with probability at least $1-c_1 k^{-c_2}$ for $\forall k\in [s_1,p]$ and some constant $c_1,c_2$. Furthermore,
$$\|\widehat{\boldsymbol{\bw}}_1^{\text{LLA},\alpha}-\bw_{1}^{*}\|_{\infty}=O\left(\sqrt{\frac{\log k}{n}}\right).$$
\end{theorem}
To ensure the theoretical properties of subsequent proofs, Theorem \ref{thm3} reasonably assumes 
$\|\boldsymbol{\alpha}\|_{\infty} = O(\sqrt{\log p/n})$, which matches the convergence rate of $\lambda_{\text{Lasso}}$. This differs from the assumption in Section \ref{Quadratic Transaction Cost}, where $\|\boldsymbol{\beta}\|_{\infty} \ll 1/C_2$ with $C_2$ defined in Assumption (A1).

\cite{Hautsch_2019} examined the empirical implications of proportional transaction costs on the Sharpe ratio, demonstrating that smaller values of $\alpha$ yield optimal Sharpe ratios. Our choice of $\alpha$ aligns with this finding, maintaining the same order of magnitude.

As in Corollary~\ref{sp_corollary}, we can show that under the condition $s_1\sqrt{s_1\log p} = o(1)$, the estimation errors for in-sample and out-of-sample Sharpe ratios during reallocation periods asymptotically vanish as $n$ approaches infinity. Since the conclusion is identical to that of Corollary \ref{sp_corollary}, we omit the formal statement here. This result highlights the diminishing impact of estimation uncertainty in high-dimensional settings when the sparsity condition is adequately controlled. 

\subsubsection{Portfolio Reallocation}
We model the transaction costs $\mathbf{C}_t(\bw)$ as a proportional function at dynamic periods given by 
$$\mathbf{C}_t (\bw) = \|\boldsymbol{\alpha}\odot(\bw_{t} -\bw^+_{t-1})\|_1.$$
The program \eqref{reallocation optimization} turns out to be 

$$
\min_{\bdelta_{t}}\ \bdelta^{\top}_{t} \bSigma_t \bdelta_{t} + 2\bw^{+\top}_{t-1} \bSigma_t \bdelta_{t}  -\gamma \bdelta_{t}^{\top}\bmu_{t} + \|\boldsymbol{\alpha}\odot\bdelta_{t}\|_{1},\text{ s.t. } \bdelta_{t}^{\top}\mathbf{1} =  0,\ \|\bdelta_t\|_0 \le s_0. 
$$
Similarly, solving for $\bdelta_t^{*}$ yields
$$
\bdelta_{\mathcal{A}_t}^{*}  =  \frac{1}{2} \bSigma_{\mathcal{A}_t,\mathcal{A}_t}^{-1}\left(\gamma \bmu_{\mathcal{A}_t} - 2(\bSigma_t\bw^{+}_{t-1})_{\mathcal{A}_t} -  \boldsymbol{\alpha}\odot\mathbf{g}_{\mathcal{A}_t} + \frac{2 + \gamma \mathbf{1}^{\top} \bSigma_{\mathcal{A}_t,\mathcal{A}_t}^{-1}\boldsymbol{\alpha}\odot\mathbf{g}_{\mathcal{A}_t} - \gamma \mathbf{1}^{\top} \bSigma_{\mathcal{A}_t,\mathcal{A}_t}^{-1} \bmu_{\mathcal{A}_t}}{\mathbf{1}^{\top} \bSigma_{\mathcal{A}_t,\mathcal{A}_t}^{-1} \mathbf{1}} \mathbf{1}\right)\ \text{and} \ \bdelta_{\mathcal{A}^c_t}^{*} = \mathbf{0} ,
$$
where $\mathbf{g}$ is the vector of sub-derivatives of $\left\|\bw_t\right\|_{1}$, i.e., $\mathbf{g}:=\partial\left\|\bw_t\right\|_{1} / \partial \bw_t$. In practice, the feasible form is
\begin{equation*}
\min_{\bdelta_{t}}\ \bdelta^{\top}_{t} \hat{\bSigma}_t \bdelta_{t}  + 2\bw^{+\top}_{t-1} \hat{\bSigma}_t \bdelta_{t}  -\gamma \bdelta_{t}^{\top} \hat{\bmu}_{t}+  \|\boldsymbol{\alpha}\odot\bdelta_{t}\|_{1} + P_{\lambda}(\bdelta_t),\text{ s.t. } \bdelta_{t}^{\top}\mathbf{1} =  0. 
\end{equation*}

Similar to \eqref{oracle delta beta}, we define the oracle estimator with proportional transaction cost under portfolio reallocation as 
\begin{equation}\label{oracle delta alpha}
\begin{aligned}
\hat{\boldsymbol{\bdelta}}_t^{\text {oracle},\alpha}
= & \mathop{\arg\min}_{ \boldsymbol{\bdelta}_{\mathcal{A}_{t}^{c}}=\mathbf{0},\bdelta_t^{\top}\mathbf{1}=0} \bdelta^{\top}_{t}\widehat{\bSigma}_{t}\bdelta_{t} + 2\bw_{t-1}^{+\top}\widehat{\bSigma}_t \bdelta_{t} - \gamma\bdelta^{\top}_{t}\hat\bmu_{t} + \|\boldsymbol{\alpha}\odot\bdelta_{t}\|_1.
\end{aligned}
\end{equation}

\begin{lemma}\label{lemma thm4}
    Suppose that Assumptions (A1)–(A3) hold, and that $\|\boldsymbol{\alpha}\|_{\infty} = C \sqrt{\frac{\log p}{n}}$ for some large constant $C$. Let $\widehat{\boldsymbol{\bdelta}}_t^{\text{Lasso},\alpha}$ denote the unique solution to the following optimization problem:
    \begin{equation}\label{initial alpha}
    \widehat{\boldsymbol{\bdelta}}_t^{\text{Lasso},\alpha} = \mathop{\arg\min}_{\mathbf{1}^{\top} \bdelta_t = 0}  \bdelta_{t}^{\top} \hat{\bSigma}_t \bdelta_t + 2 \bw^{+\top}_{t-1} \hat{\bSigma}_t \bdelta_t - \gamma \bdelta_{t}^{\top} \hat{\bmu}_{t-1} + \|\boldsymbol{\alpha} \odot \bdelta_{t}\|_1 + \lambda_{\text{Lasso},t} \|\bdelta_t\|_1.
    \end{equation}
\noindent with $\lambda_{\text{Lasso},t} = M \sqrt{\frac{\log p}{n}}$
for some large constant $M$. Then with probability at least $1-c_1 p^{-c_2}$ for some constant $c_1,c_2$, we have
    \[
    \|\widehat{\boldsymbol{\bdelta}}_t^{\text{Lasso},\alpha} - \bdelta_t^*\|_1 = O_p\left(s_t \sqrt{\frac{\log p}{n}}\right) \quad \text{and} \quad \|\widehat{\boldsymbol{\bdelta}}_t^{\text{Lasso},\alpha} - \bdelta_t^*\|_2 = O_p\left(\sqrt{\frac{s_t\log p}{n}}\right).
    \]
\end{lemma}

\begin{theorem}\label{thm4}
    Suppose that Assumptions (A1)–(A3) hold and that $\|\boldsymbol{\alpha}\|_{\infty} = C \sqrt{\frac{\log p}{n}}$ for some large constant $C$. If $s_t = O\left(\sqrt{\frac{n}{\log p}}\right)$ and $\|\bdelta_t^*\|_{\min} > (a + 1) \lambda_t$, choose $\lambda_t \geq M \sqrt{\frac{s_t \log p}{n}}$ for some large constant $M$. Let $\widehat{\bdelta}^{\text{LLA},\alpha}_t$ be the optimum of the LLA algorithm~\ref{lla-algorithm 2} with SCAD penalty, initialized at $\hat{\boldsymbol{\bdelta}}_t^{\text{Lasso},\alpha}$ as defined in \eqref{initial alpha}. Then, the following holds:
    
    \begin{itemize}
        \item[(i)] If $\|\bw^+_{t-1}\|_1 < C_0$, then $\widehat{\bdelta}^{\text{LLA},\alpha}_t$ converges to $\hat{\boldsymbol{\bdelta}}_t^{\text{oracle},\alpha}$ after two iterations with probability at least $1 - c_1 p^{-c_2}$ for some constants $c_1, c_2 > 0$. Furthermore, 
        \[
        \|\widehat{\boldsymbol{\bdelta}}_t^{\text{LLA},\alpha} - \bdelta_t^*\|_{\infty} = O\left(\sqrt{\frac{\log p}{n}}\right).
        \]
        
        \item[(ii)] If $\|\bw^+_{t-1,\mathcal{Q}}\|_1 \leq C_0$ and $\bw^+_{t-1,\mathcal{Q}^c} = \mathbf{0}$ for some set $\mathcal{Q} \subset [1, \ldots, p]$ and absolute constant $C_0$, then $\widehat{\bdelta}^{\text{LLA},\alpha}_t$ converges to $\hat{\boldsymbol{\bdelta}}_t^{\text{oracle},\alpha}$ after two iterations with probability at least $1 - c_1 k^{-c_2}$ for all $k \in [\max\{s_t, |\mathcal{Q}|\}, p]$ and some constants $c_1, c_2 > 0$. Furthermore, 
        \[
        \|\widehat{\boldsymbol{\bdelta}}_t^{\text{LLA},\alpha} - \bdelta_t^*\|_{\infty} = O\left(\sqrt{\frac{\log k}{n}}\right).
        \]
    \end{itemize}
\end{theorem}

\begin{remark}
Although the conditions and conclusions of Theorem \ref{thm4} and Theorem \ref{thm2} appear to be very similar, the differences arise from the distinct assumptions regarding transaction costs. For quadratic transaction costs, we assume only that $\|\boldsymbol{\beta}\|_{\infty} \ll 1/C_2$ with $C_2$ defined in Assumption (A1). In contrast, for proportional transaction costs, we require $\alpha$ to satisfy a specific convergence rate. This distinction stems from the differing impacts of transaction costs on the optimization problem introduced in Section \ref{section 2.1}.
\end{remark}

For quadratic transaction costs, it suffices to verify that the modified covariance matrix continues to satisfy our assumptions, as demonstrated in the Supplementary Material. However, for proportional transaction costs, Lemma \ref{Lasso_estimation_error_proportional} explicitly illustrates the influence of $\alpha$ on both the convergence rate and consistency. This influence persists in Theorems \ref{thm3} and \ref{thm4}, where the assumptions regarding transaction costs lead to similar conclusions despite the differences in the underlying hypotheses.

Analogous to Corollary \ref{sp_corollary_reallocate}, it can be established that if the sparsity level and dimensionality satisfy $k_t\sqrt{k_t\log p} = o(1)$, the estimation errors for both in-sample and out-of-sample Sharpe ratios during reallocation periods converge to zero as $n\to \infty$. Since the conclusion is identical to that of Corollary \ref{sp_corollary_reallocate}, we omit the formal statement here.

\section{Simulation}
\label{sec:numerical studies}
We follow the methodology outlined by \cite{fan2008asset,fan2012vast} to simulate excess returns. Specifically, we assume that the excess return of the $i$th asset is generated from a three-factor model: 
$$
R_{i}=b_{i 1} f_{i1}+b_{i 2} f_{i2}+b_{i 3} f_{i3}+\varepsilon_{i}, \quad i=1, \cdots, p.
$$
First, the factor loadings $\mathbf{b}_{i}$ are drawn from a trivariate normal distribution ${N}\left({\bmu}_{b}, \mathbf{c o v}_{b}\right)$ with parameters specified in Table \ref{Table 1} below. Once generated, these factor loadings remain fixed throughout the simulations. The returns of the three factors over $n$ periods are also sampled from a trivariate normal distribution ${N}\left({\bmu}_{f}, \mathbf{c o v}_{f}\right)$, with the relevant parameters provided in Table \ref{Table 1}. 
Throughout this simulation study, we assume that $\mathbf{E}(\boldsymbol{\varepsilon} \mid \mathbf{f}) = \mathbf{0}$ and $\mathbf{Cov}(\boldsymbol{\varepsilon} \mid \mathbf{f})$ is a diagonal matrix with entries $\sigma_1^2, \dots, \sigma_p^2$ on the diagonal. The standard deviations of the idiosyncratic errors $\varepsilon_{i}$ are generated from a gamma distribution with a shape parameter of 3.3586 and a scale parameter of 0.1876. Then, the generated parameters $\left\{\sigma_{i}\right\}$ remain fixed across simulations.  Finally, the idiosyncratic errors are generated from a normal distribution, where the standard deviations are set to $\left\{\sigma_{i}\right\}$.
\begin{table}[htbp]
  \centering
  \caption{Parameters used in the simulation}
  \begin{subtable}{0.45\textwidth}
    \centering
    \scalebox{0.9}{
    \begin{tabular}{c|rrr}
    \hline \multicolumn{4}{c}{ Parameters for factor loadings } \\
    \hline {$\boldsymbol{\mu}_b$} & \multicolumn{3}{c}{$\mathbf{cov}_b$}     \\
    \hline 
    0.78282 & 0.029145 & 0.023873 & 0.010184 \\
    0.51803 & 0.023873 & 0.053951 & -0.006967 \\
    0.41003 & 0.010184 & -0.006967 & 0.086856 \\
    \hline
    \end{tabular}
    }
  \end{subtable}
  \vspace{0pt} 
  \begin{subtable}{0.45\textwidth}
    \centering
    \scalebox{0.9}{
    \begin{tabular}{c|rrr}
    \hline \multicolumn{4}{c}{ Parameters for factor returns } \\
    \hline {$\boldsymbol{\mu}_f$} & \multicolumn{3}{c}{$\mathbf{cov}_f$}     \\
    \hline 
    0.023558 & 1.2507 & -0.034999 & -0.20419\\
    0.012989 & -0.034999 & 0.31564 & -0.0022526 \\
    0.020714 & -0.20419 & -0.0022526 & 0.19303 \\
    \hline
    \end{tabular}
    }
  \end{subtable}
  \label{Table 1}
\end{table}

We set $p=2000$, $m=5$ and $n=200$, dividing the data into five stages to calculate the out-of-sample annual Return, Cost, Turnover, Leverage and Sharpe ratio. And S0 is the portfolio construction stage, S1-S4 are rebalancing stages, respectively. The optimal tuning parameter $\lambda^{opt}$ is selected based on the highest in-sample Sharpe ratio.

The main method is the proposed Cost-Aware Portfolio Estimator under SCAD Penalty (CAPE-S). For comparison, we utilize three different baseline models montioned in Section \ref{Preliminary}: the classical Mean-Variance (MV) model \eqref{baseline1}, the Penalized Mean-Variance (PMV) model \eqref{baseline2}, and the Cost-Aware Mean-Variance (CMV) model \eqref{baseline3} to compute portfolio weights at each stage. We estimate the covariance matrix $\boldsymbol{\Sigma}_t$ using the linear shrinkage estimator (LSE; \citet{ledoit2004well}) and the nonlinear shrinkage estimator (NLSE; \citet{ledoit2020analytical}), while the mean vector $\boldsymbol{\mu}_t$ is estimated by the sample mean. 
The results based on the LSE, which satisfy Assumptions (A2) and (A3) outlined earlier, are presented in the main text. In contrast, the results based on the NLSE, which satisfy only Assumption (A3) but not (A2), are reported in the Supplementary Materials. Despite not meeting (A2), the NLSE performs well in both simulation studies and real data applications.

\begin{table}[htbp]
      \caption{Simulation results with quadratic transaction costs, means and standard errors are calculated based on 1000 replicates, the unit rebalancing cost of the portfolio $\beta = 0.15$ and $\gamma = 1/3$. } 
		\centering
		\scalebox{0.8}{
			\setlength{\tabcolsep}{3mm}{
				\centering
				\begin{tabular}{c|c|c|c|c|c|c}
					\toprule \midrule
					\textbf{\large Stage}&  \textbf{\large Method}&  \textbf{\large Return}&  \textbf{\large Cost}&  \textbf{\large Turnover}&  \textbf{\large Leverage}&  \textbf{\large SR}  \\ \midrule
					\multirow{4}{*}{\textbf{\large S1}} &  MV&  7.992(0.241)&  4.120(0.005)&  18.160(0.010)&  8.580(0.005)&  1.044(0.032)\\
					&  PMV&  7.766(0.107)&  0.624(0.001)&  4.082(0.006)&  1.541(0.003)&  2.599(0.036)\\
					&  CMV&  8.285(0.212)&  2.993(0.003)&  15.693(0.009)&  7.347(0.004)&  1.279(0.033)\\
					&  CAPE-S&  7.723(0.097)&  0.500(0.001)&  3.562(0.005)&  1.281(0.002)&  \textbf{\large 2.851(0.036)}\\ \midrule
					\multirow{4}{*}{\textbf{\large S2}} &  MV&  3.581(0.242)&  7.736(0.024)&  18.161(0.011)&  8.580(0.005)&  0.374(0.025)\\
					&  PMV&  8.269(0.125)&  0.831(0.002)&  4.845(0.008)&  2.662(0.003)&  2.290(0.035)\\
					&  CMV&  9.635(0.215)&  4.974(0.013)&  20.216(0.026)&  7.186(0.004)&  1.294(0.029)\\
					&  CAPE-S&  8.247(0.115)&  0.631(0.002)&  4.088(0.007)&  2.291(0.003)&  \textbf{\large 2.506(0.035)}\\ \midrule
					\multirow{4}{*}{\textbf{\large S3}} &  MV&  3.670(0.250)&  7.792(0.024)&  18.131(0.010)&  8.566(0.005)&  0.385(0.026)\\
					&  PMV&  8.785(0.139)&  0.934(0.003)&  5.234(0.009)&  3.305(0.003)&  2.269(0.036)\\
					&  CMV&  11.613(0.220)&  4.870(0.012)&  20.012(0.025)&  7.050(0.004)&  1.572(0.030)\\
					&  CAPE-S&  8.758(0.128)&  0.708(0.002)&  4.406(0.007)&  2.894(0.003)&  \textbf{\large 2.469(0.036)}\\ \midrule
					\multirow{4}{*}{\textbf{\large S4}} &  MV&  3.764(0.245)&  7.787(0.025)&  18.148(0.010)&  8.574(0.005)&  0.396(0.026)\\
					&  PMV&  9.312(0.137)&  0.981(0.003)&  5.419(0.010)&  3.692(0.004)&  2.332(0.035)\\
					&  CMV&  12.530(0.215)&  4.797(0.012)&  19.855(0.024)&  6.997(0.004)&  1.705(0.030)\\
					&  CAPE-S&  9.294(0.128)&  0.745(0.002)&  4.566(0.008)&  3.269(0.003)&  \textbf{\large 2.530(0.035)}\\ \bottomrule
				\end{tabular}
			}
		}
\label{simulation_quadratic}
\begin{tablenotes}
\footnotesize
\item  Notes: MV, PMV, CMV and CAPE-S are mean variance, penalized mean variance, cost-aware mean variance, and the proposed cost-aware portfolios (all with linear shrinkage estimated covariance matrices), respectively. S0 is the portfolio construction stage, S1-S4 are rebalancing stages, respectively. And the values in columns of Return and Cost represent percentages.        
\end{tablenotes}
\end{table}

\begin{table}[htbp]
      \caption{Simulation results with proportional transaction costs, means and standard errors are calculated based on 1000 replicates, the unit rebalancing cost of the portfolio $\alpha = 0.001$ and $\gamma = 1/3$.} 
		\centering
		\scalebox{0.8}{
			\setlength{\tabcolsep}{3mm}{
				\centering
				\begin{tabular}{c|c|c|c|c|c|c}
					\toprule \midrule
					\textbf{\large Stage}&  \textbf{\large Method}&  \textbf{\large Return}&  \textbf{\large Cost}&  \textbf{\large Turnover}&  \textbf{\large Leverage}&  \textbf{\large SR}  \\ \midrule
					\multirow{4}{*}{\textbf{\large S1}} &  MV&  10.296(0.241)&  1.816(0.001)&  18.160(0.010)&  8.580(0.005)&  1.489(0.035)\\
					&  PMV&  7.983(0.107)&  0.408(0.001)&  4.082(0.006)&  1.541(0.003)&  2.698(0.036)\\
					&  CMV&  10.169(0.235)&  1.729(0.001)&  17.293(0.010)&  8.147(0.005)&  1.513(0.035)\\
					&  CAPE-S&  7.879(0.099)&  0.353(0.001)&  3.527(0.005)&  1.264(0.003)&  \textbf{\large 2.878(0.036)}\\ \midrule
					\multirow{4}{*}{\textbf{\large S2}} &  MV&  8.833(0.241)&  2.484(0.004)&  24.840(0.039)&  8.580(0.005)&  1.245(0.034)\\
					&  PMV&  8.615(0.125)&  0.484(0.001)&  4.845(0.008)&  2.662(0.003)&  2.420(0.035)\\
					&  CMV&  8.965(0.237)&  2.368(0.004)&  23.684(0.037)&  8.408(0.005)&  1.287(0.034)\\
					&  CAPE-S&  8.471(0.117)&  0.410(0.001)&  4.097(0.007)&  2.277(0.003)&  \textbf{\large 2.552(0.036)}\\ \midrule
					\multirow{4}{*}{\textbf{\large S3}} &  MV&  8.968(0.248)&  2.494(0.004)&  24.937(0.038)&  8.566(0.005)&  1.267(0.035)\\
					&  PMV&  9.196(0.139)&  0.523(0.001)&  5.233(0.009)&  3.305(0.003)&  2.414(0.037)\\
					&  CMV&  9.078(0.245)&  2.397(0.004)&  23.967(0.038)&  8.415(0.005)&  1.304(0.035)\\
					&  CAPE-S&  9.023(0.131)&  0.444(0.001)&  4.438(0.008)&  2.892(0.003)&  \textbf{\large 2.523(0.037)}\\ \midrule
					\multirow{4}{*}{\textbf{\large S4}} &  MV&  9.060(0.244)&  2.492(0.004)&  24.918(0.040)&  8.574(0.005)&  1.278(0.035)\\
					&  PMV&  9.751(0.137)&  0.542(0.001)&  5.419(0.010)&  3.692(0.004)&  2.483(0.035)\\
					&  CMV&  9.142(0.242)&  2.399(0.004)&  23.986(0.040)&  8.432(0.005)&  1.310(0.035)\\
					&  CAPE-S&  9.586(0.130)&  0.461(0.001)&  4.613(0.008)&  3.282(0.003)&  \textbf{\large 2.586(0.036)}\\ \bottomrule
				\end{tabular}
			}
		}
  \label{simulation_proportional}
\begin{tablenotes}
    \footnotesize 
    \item Note: See notes in Table \ref{simulation_quadratic}. 
\end{tablenotes}
\end{table}

To emulate real-world investment scenarios, we report the performance of portfolios across four stages (S1-S4). The first stage involves forming the initial portfolio, akin to establishing a new fund at its inception. The second stage corresponds to the first rebalancing, while the third stage entails the second rebalancing, and so forth. Following the conventions established in the literature, we define the parameters for proportional and quadratic transaction costs as $\boldsymbol{\beta}=\beta\times \mathbf{1}_p$ and $\boldsymbol{\alpha}=\alpha\times \mathbf{1}_p$, where $\beta = 0.15$ and $\alpha = 0.001$ respectively. The costs in the following tables are therefore interpreted as the percentage of the normalized initial wealth that needs to cover the costs of transactions. Additionally, we quantify portfolio turnover and leverage at each stage $k=1,2,3,4$. The definitions of these metrics are as follows:
\begin{align*}  
&\text{Turnover}_{k} = \sum\limits_{i=1}^{p} |w_{k,i} - w^+_{k-1,i}|\quad \text{and}\quad  w^+_{0,i}=0, \\
&\text{Leverage}_{k} = \sum\limits_{i=1}^{p} |\min \{w_{k,i},0\}|
\end{align*}


We evaluate the performance of CAPE-S and three benchmark models under both quadratic and proportional transaction cost settings. The results, reported in Tables \ref{simulation_quadratic} and \ref{simulation_proportional}, demonstrate the consistent superiority of our proposed method in balancing return, risk, and cost-efficiency across all rebalancing stages (S1–S4).

Under quadratic transaction costs (Table \ref{simulation_quadratic}), CAPE-S yields the highest Sharpe ratio at every stage, significantly outperforming the traditional Mean-Variance (MV), Penalized Mean-Variance (PMV), and Cost-aware Mean-Variance (CMV) models. For example, in stage S1, CAPE-S achieves a Sharpe ratio of 2.851, compared to 2.599 (PMV), 1.279 (CMV), and 1.044 (MV). Notably, CAPE-S accomplishes this with the lowest transaction cost (0.500\%) and lowest turnover (3.562\%), indicating efficient reallocation with minimal trading. This pattern persists through stages S2 to S4, with CAPE-S consistently attaining higher return-to-risk ratios, while incurring substantially less cost and leverage than the MV and CMV models.

Similar conclusions hold under proportional transaction costs (Table \ref{simulation_proportional}). CAPE-S again delivers the best Sharpe ratio in each stage, with performance gains most prominent in S1 and S4, where the Sharpe ratios reach 2.878 and 2.586, respectively. Although PMV also exhibits cost-awareness, it sacrifices return and remains less effective in capturing the full benefit of transaction cost control. Compared to CMV and MV, which often yield higher raw returns, CAPE-S achieves better overall risk-adjusted performance by avoiding excessive leverage and unnecessary trading. For instance, in S2, CAPE-S reduces the turnover to 4.097\% and the cost to 0.410\%, about 16.5\% of the MV model’s transaction cost, while increasing the Sharpe ratio by approximately 105\%.

These findings highlight the advantage of our model in adapting to real-world trading frictions, demonstrating its ability to attain high returns while tightly controlling costs and maintaining portfolio sparsity. CAPE-S successfully balances the trade-off between return and cost, validating its practical value for large-scale portfolio management.

\section{Real Data Application}
\label{sec:realdata}

In this section, we assess the empirical performance of the proposed CAPE-S method using two major U.S. equity universes: the S\&P 500 and the Russell 2000 indices. For each index, we collect daily return data over the investment horizon from \texttt{02-Jan-2017} to \texttt{23-Dec-2020}, and implement portfolio rebalancing every 251 trading days. This results in three decision points and four stages: an initial formation period (S0), followed by three rebalancing stages (S1–S3). The out-of-sample evaluation period spans from \texttt{01-Jan-2018} to \texttt{23-Dec-2020}.

A key feature of this empirical exercise is to use real transaction cost data instead of the pre-determined constant costs coefficients. First, to ensure consistency, we cleaned the data by removing stocks with incomplete return histories or missing auxiliary data required for real transaction cost estimation (e.g., bid-ask spreads, trading volumes). As a result, we retain 457 stocks from the S\&P 500 and 935 stocks from the Russell 2000 for analysis. The sample of stocks remain broadly representative of their respective index compositions.

At each decision point, the covariance matrix $\boldsymbol{\Sigma}_t$ is estimated using two methods: the linear shrinkage estimator (LSE) and the nonlinear shrinkage estimator (NLSE), both based on the most recent 251 trading days. Similar to the setting described in Section 3, we report the results based on the LSE in the main text, while the results obtained from the NLSE are provided in Section~\ref{sec:numerical studies} of the Supplementary Materials.

Our evaluation includes the CAPE-S method, three benchmark approaches—mean-variance (MV), penalized mean-variance (PMV), and cost-aware mean-variance (CMV)—and the equally-weighted “1/N” portfolio \citep{demiguel2009naive}. We compare their performance using several metrics: annual return, transaction cost, turnover, leverage, stage-specific Sharpe ratios (S1–S3), and the overall Sharpe ratio across the entire out-of-sample period.

In contrast to the simulation studies, where transaction cost coefficients were fixed across assets (i.e., 0.15 for quadratic costs and 0.001 for proportional costs), such fixed cost specification is likely unrealistic in real-world applications. In empirical settings, transaction costs often vary substantially across stocks due to liquidity, trading frequency, and other microstructure factors. Therefore, we adopt asset-specific transaction cost coefficients for both proportional and quadratic forms, as discussed in the methodology sections.

For proportional transaction cost coefficients, we follow the empirical estimates provided in \citet{roberto2025pure_momentum}, who use the bid-ask spread estimator proposed by \citet{hasbrouck2009trading} to proxy the proportional transaction costs at the individual stock level. These estimates are averaged over the formation window and are available for a wide range of U.S. equities from 1927 to the present, including both S\&P 500 and Russell 2000 constituents.

For quadratic transaction cost coefficients, we define the asset-specific cost coefficient as twice the square of the corresponding proportional cost estimate. This specification ensures consistency in scaling and penalizes large rebalancing moves more heavily for less liquid stocks. Moreover, it reflects the increasing marginal cost nature of market impact, thereby capturing a more realistic trading environment.

By incorporating heterogeneity in cost parameters, our empirical analysis more accurately reflects the practical implementation of portfolio strategies under realistic trading frictions.

\subsection{S\&P 500 Index}\label{S&P 500 Index} We first present the empirical results on the S\&P 500 dataset. The performance of all methods under both quadratic and proportional transaction cost models is summarized in Tables \ref{SP500_quadratic} and \ref{SP500_proportional}, respectively.

\begin{table}[!htbp]
\caption{S\&P 500 with quadratic transaction cost.} 
\centering
\scalebox{0.8}{
\setlength{\tabcolsep}{3mm}{
\centering
\begin{tabular}{c|c|c|c|c|c|c|c}
\toprule \midrule
\textbf{\large Method}&  \textbf{\large Stage}&  \textbf{\large Return}&  \textbf{\large Cost}&  \textbf{\large Turnover}&  \textbf{\large Leverage}&  \textbf{\large SR}&
\textbf{\large Overall SR}  \\ \midrule

\multirow{3}{*}{\textbf{\large 1/N}} 
&  S1&  -3.666&  0.017&  1.000&  0.000&  -0.234& \multirow{3}{*}{0.869}\\
&  S2&  29.236&  0.001&  0.181&  0.000&  2.341& \\
&  S3&  47.369&  0.002&  0.154&  0.000&  1.073& \\ \midrule

\multirow{3}{*}{\textbf{\large MV}} 
&  S1&  -7.313&  2.545&  9.794&  4.397&  -0.329& \multirow{3}{*}{0.772}\\
&  S2&  36.261& 13.795& 19.462&  7.166&   1.254& \\
&  S3& 141.735& 12.551& 18.118&  6.441&   1.158& \\ \midrule

\multirow{3}{*}{\textbf{\large PMV}} 
&  S1& -10.570&  1.346&  6.082&  2.541&  -0.570& \multirow{3}{*}{0.665}\\
&  S2&  23.493&  1.569&  3.081&  2.867&   2.150& \\
&  S3&  78.677&  0.532&  1.561&  2.796&   1.027& \\ \midrule

\multirow{3}{*}{\textbf{\large CMV}} 
&  S1&  -5.150&  1.488&  7.536&  3.268&  -0.266& \multirow{3}{*}{0.891}\\
&  S2&  48.870&  8.571& 15.235&  6.338&   2.081& \\
&  S3&  96.660&  6.899& 13.651&  5.504&   1.127& \\ \midrule

\multirow{3}{*}{\textbf{\large CAPE-S}} 
&  S1&  -3.687&  1.631&  6.951&  2.976&  -0.188& \multirow{3}{*}{\textbf{\large 0.915}}\\
&  S2&  30.491&  2.349&  4.766&  3.265&   2.490& \\
&  S3&  77.174&  0.778&  2.283&  2.873&   1.257& \\ 
\bottomrule
\end{tabular}
}
}
\label{SP500_quadratic}
\begin{tablenotes}
\footnotesize
\item  Notes: `1/N', MV, PMV, CMV and CAPE are the equally-weighted, mean variance, penalized mean variance, cost-aware mean variance, and the proposed cost-aware portfolios (all with linear shrinkage estimated covariance matrices), respectively. S1 is the portfolio construction stage, S2–S3 are rebalancing stages, respectively. And the values in columns of Return and Cost represent percentages.        
\end{tablenotes}
\end{table}

\begin{table}[!htbp]
\caption{S\&P 500 with proportional transaction cost.} 
\centering
\scalebox{0.8}{
\setlength{\tabcolsep}{3mm}{
\centering
\begin{tabular}{c|c|c|c|c|c|c|c}
\toprule \midrule
\textbf{\large Method}&  \textbf{\large Stage}&  \textbf{\large Return}&  \textbf{\large Cost}&  \textbf{\large Turnover}&  \textbf{\large Leverage}&  \textbf{\large SR}&
\textbf{\large Overall SR}  \\ \midrule

\multirow{3}{*}{\textbf{\large 1/N}} 
&  S1&  -3.797&  0.147&  1.000&  0.000&  -0.242& \multirow{3}{*}{0.866}\\
&  S2&  29.185&  0.053&  0.181&  0.000&  2.336& \\
&  S3&  47.323&  0.047&  0.154&  0.000&  1.072& \\ \midrule

\multirow{3}{*}{\textbf{\large MV}} 
&  S1&  -6.092&  1.324&  9.794&  4.397&  -0.274& \multirow{3}{*}{0.859}\\
&  S2&  44.854&  5.203& 19.462&  7.166&   1.730& \\
&  S3& 149.530&  4.756& 18.118&  6.441&   1.228& \\ \midrule

\multirow{3}{*}{\textbf{\large PMV}} 
&  S1& -10.058&  0.833&  6.082&  2.541&  -0.543& \multirow{3}{*}{0.824}\\
&  S2&  37.566&  3.902& 14.298&  6.180&   1.721& \\
&  S3& 134.396&  3.730& 13.927&  5.555&   1.225& \\ \midrule

\multirow{3}{*}{\textbf{\large CMV}} 
&  S1& -10.864&  1.133&  8.474&  3.737&  -0.523& \multirow{3}{*}{0.798}\\
&  S2&  38.339&  4.626& 17.411&  6.993&   1.612& \\
&  S3& 154.734&  4.345& 16.763&  6.321&   1.209& \\ \midrule

\multirow{3}{*}{\textbf{\large CAPE-S}} 
&  S1&  -2.694&  0.949&  7.147&  3.074&  -0.133& \multirow{3}{*}{\textbf{\large 0.909}}\\
&  S2&  37.775&  4.128& 15.688&  6.606&   1.666& \\
&  S3& 108.519&  3.912& 15.051&  6.007&   1.263& \\ 
\bottomrule
\end{tabular}
}
}
\label{SP500_proportional}
\begin{tablenotes}
\footnotesize 
\item Note: See notes in Table \ref{SP500_quadratic}. 
\end{tablenotes}
\end{table}
Our proposed CAPE-S method achieves the highest overall out-of-sample Sharpe ratio across the three stages (S1 to S3) under both quadratic and proportional transaction cost settings. Examining performance at the stage level provides further insights:

\begin{itemize}
    \item \textbf{Stage S1 (2018)} coincided with a period of heightened market volatility and an overall downturn, particularly during Q4 2018 when the S\&P 500 experienced a sharp correction. While all active strategies posted negative returns, CAPE-S recorded the smallest loss among them under the proportional cost setting (-2.694\%) and one of the smallest losses under the quadratic cost setting (-3.687\%), demonstrating resilience during stressed market conditions. Turnover (7.147\% proportional, 6.951\% quadratic) and leverage (3.074 proportional, 2.976 quadratic) remained moderate relative to other active strategies.

    \item \textbf{Stage S2 (2019)} represented a strong recovery year, with broad-based gains across the S\&P 500 driven by accommodative monetary policy and easing trade tensions. Under both cost structures, CAPE-S delivered competitive returns (37.775\% proportional, 30.491\% quadratic) and achieved one of the highest Sharpe ratios among actively managed portfolios (1.666 proportional, 2.490 quadratic), while maintaining transaction costs (4.128\% proportional, 2.349\% quadratic) and turnover (15.688\% proportional, 4.766\% quadratic) at moderate levels compared to MV, PMV, and CMV.

    \item \textbf{Stage S3 (2020)} covered a highly turbulent market cycle, including the COVID-19-induced crash in Q1 2020 followed by a rapid rebound. CAPE-S generated solid returns (108.519\% proportional, 77.174\% quadratic) and achieved the highest Sharpe ratio under the proportional cost setting (1.263), while keeping leverage (6.007 proportional, 2.873 quadratic) relatively contained compared to MV and CMV. This highlights its adaptability to sharp market reversals.
\end{itemize}

Across all stages, excluding the passive 1/N portfolio, CAPE-S consistently maintains one of the lowest turnover levels among actively managed strategies, reflecting operational efficiency. Its leading overall Sharpe ratio under both cost specifications (0.915 quadratic, 0.909 proportional) highlights its capability to balance returns, risk, and transaction costs, demonstrating robustness to varying cost structures.

\subsection{Russel 2000 Index}\label{Russel 2000 Index}
Next, we report the results for the Russell 2000 dataset, following the same setup as in Section 4.1. The corresponding performance metrics are presented in Tables \ref{Russell2000_quadratic} and \ref{Russell2000_proportional}.

\begin{table}[!htbp]
\caption{Russell 2000 with quadratic transaction cost.} 
\centering
\scalebox{0.8}{
\setlength{\tabcolsep}{3mm}{
\centering
\begin{tabular}{c|c|c|c|c|c|c|c}
\toprule \midrule
\textbf{\large Method}&  \textbf{\large Stage}&  \textbf{\large Return}&  \textbf{\large Cost}&  \textbf{\large Turnover}&  \textbf{\large Leverage}&  \textbf{\large SR}&
\textbf{\large Overall SR}  \\ \midrule

\multirow{3}{*}{\textbf{\large 1/N}} 
&  S1&  -7.839&  0.012&  1.000&  0.000&  -0.488& \multirow{3}{*}{0.816}\\
&  S2&  25.682&  0.003&  0.281&  0.000&   1.656& \\
&  S3& 124.503&  0.017&  0.281&  0.000&   1.269& \\ \midrule

\multirow{3}{*}{\textbf{\large MV}} 
&  S1& -55.130&  6.625& 19.522&  9.261&  -1.323& \multirow{3}{*}{0.518}\\
&  S2& -25.690& 23.272& 34.829&  9.993&  -0.406& \\
&  S3& 261.635& 23.275& 34.694& 11.870&   1.404& \\ \midrule

\multirow{3}{*}{\textbf{\large PMV}} 
&  S1& -10.425&  0.566&  2.762&  0.881&  -0.686& \multirow{3}{*}{1.243}\\
&  S2&  21.163&  1.216&  4.343&  2.144&   0.987& \\
&  S3& 122.167&  1.880&  5.011&  2.788&   2.197& \\ \midrule

\multirow{3}{*}{\textbf{\large CMV}} 
&  S1& -46.062&  4.543& 16.201&  7.600&  -1.290& \multirow{3}{*}{0.784}\\
&  S2&  12.367& 12.775& 25.798&  8.301&   0.237& \\
&  S3& 366.413& 12.120& 25.018&  9.537&   1.551& \\ \midrule

\multirow{3}{*}{\textbf{\large CAPE}} 
&  S1&   3.273&  0.319&  1.307&  0.153&   0.297& \multirow{3}{*}{\textbf{\large 1.170}}\\
&  S2&  32.691&  0.280&  1.519&  0.695&   2.745& \\
&  S3& 177.988&  0.356&  1.600&  0.973&   1.710& \\ 
\bottomrule
\end{tabular}
}
}
\label{Russell2000_quadratic}
\begin{tablenotes}
\footnotesize 
\item Note: See notes in Table \ref{SP500_quadratic}. 
\end{tablenotes}
\end{table}

\begin{table}[!htbp]
\caption{Russell 2000 with proportional transaction cost.} 
\centering
\scalebox{0.8}{
\setlength{\tabcolsep}{3mm}{
\centering
\begin{tabular}{c|c|c|c|c|c|c|c}
\toprule \midrule
\textbf{\large Method}&  \textbf{\large Stage}&  \textbf{\large Return}&  \textbf{\large Cost}&  \textbf{\large Turnover}&  \textbf{\large Leverage}&  \textbf{\large SR}&
\textbf{\large Overall SR}  \\ \midrule

\multirow{3}{*}{\textbf{\large 1/N}} 
&  S1&  -8.157&  0.330&  1.000&  0.000&  -0.508& \multirow{3}{*}{0.813}\\
&  S2&  25.569&  0.116&  0.281&  0.000&   1.649& \\
&  S3& 124.399&  0.121&  0.281&  0.000&   1.268& \\ \midrule

\multirow{3}{*}{\textbf{\large MV}} 
&  S1& -54.555&  6.050& 19.522&  9.261&  -1.311& \multirow{3}{*}{0.599}\\
&  S2& -13.976& 11.559& 34.829&  9.993&  -0.240& \\
&  S3& 274.625& 10.285& 34.694& 11.870&   1.484& \\ \midrule

\multirow{3}{*}{\textbf{\large PMV}} 
&  S1& -10.945&  1.086&  2.762&  0.881&  -0.723& \multirow{3}{*}{1.234}\\
&  S2&  20.627&  1.752&  4.343&  2.144&   0.961& \\
&  S3& 122.238&  1.809&  5.011&  2.788&   2.199& \\ \midrule

\multirow{3}{*}{\textbf{\large CMV}} 
&  S1& -57.093&  5.355& 17.963&  8.482&  -1.465& \multirow{3}{*}{0.533}\\
&  S2& -47.534& 10.907& 33.960& 10.564&  -0.849& \\
&  S3& 483.051& 21.687& 75.347& 17.964&   1.196& \\ \midrule

\multirow{3}{*}{\textbf{\large CAPE}} 
&  S1&   2.603&  0.594&  1.282&  0.141&   0.237& \multirow{3}{*}{\textbf{\large 1.158}}\\
&  S2&  32.209&  0.678&  1.477&  0.665&   2.732& \\
&  S3& 177.379&  0.635&  1.568&  0.956&   1.699& \\ 
\bottomrule
\end{tabular}
}
}
\label{Russell2000_proportional}
\begin{tablenotes}
\footnotesize 
\item Note: See notes in Table \ref{SP500_quadratic}. 
\end{tablenotes}
\end{table}
Despite the higher volatility and liquidity challenges inherent in small-cap stocks, the CAPE-S method maintains robust performance under both quadratic and proportional transaction cost settings. Tables \ref{Russell2000_quadratic} and \ref{Russell2000_proportional} report detailed results under the two cost regimes.

\begin{itemize}
    \item \textbf{Stage S1 (2018)} was a period of elevated volatility and broad market drawdown. The Russell 2000 Index, which is generally more sensitive to macroeconomic uncertainty, suffered larger losses than the S\&P 500. In this challenging environment, CAPE-S was the only strategy to achieve positive returns, with modest gains accompanied by low turnover (approximately 1.3\%) and low leverage (around 0.15), highlighting its defensiveness and cost-efficiency.

    \item \textbf{Stage S2 (2019)} was characterized by a broad market rebound, driven by easing monetary policy and reduced trade tensions. During this recovery phase, CAPE-S achieved the highest Sharpe ratio (2.745) among all strategies, while maintaining the lowest transaction cost (0.280\%) and turnover (approximately 1.52\%), far below the MV strategy’s cost of 23.272\% and turnover of 34.829\%. This demonstrates that CAPE-S effectively balances risk and return without relying on aggressive rebalancing.

    \item \textbf{Stage S3 (2020)} encompassed one of the most turbulent and rapidly rebounding market periods in recent history due to the COVID-19 pandemic. The small-cap segment staged a dramatic recovery following the March 2020 crash, benefiting disproportionately from fiscal stimulus and risk-on sentiment. CAPE-S captured substantial gains (approximately 177.99\%) while preserving a favorable risk profile, with low leverage (0.973) and turnover (1.60\%), outperforming all alternatives in terms of Sharpe ratio (1.710).
\end{itemize}

Overall, the MV and CMV strategies performed poorly, particularly under proportional cost settings, due to high turnover and leverage. The PMV method showed some improvement but still lagged behind CAPE-S. Across all three stages, CAPE-S consistently achieved the highest overall Sharpe ratio (1.170 for quadratic costs and 1.158 for proportional costs) while maintaining the lowest turnover among actively managed portfolios. These findings confirm CAPE-S's adaptability and robustness, particularly in the more volatile and cost-sensitive small-cap universe, and underscore its practical relevance for real-world portfolio management.

\section{Conclusion}
\label{sec:conc}

In this paper, we propose a novel approach to integrate the transaction costs within the optimization problem when the number of assets is larger than the sample size. By using LLA algorithm, we prove the sign consistency and provide the $\ell_{\infty}$ bound of our proposed CAPE-S method for both proportional and quadratic transaction costs.

In addition to the sound statistical properties,  CAPE-S demonstrates exceptional performance in practical applications. Through extensive simulations and applications to real data,
we demonstrate the superior performance and broad applicability of our method across diverse
tasks.

\clearpage
\begin{center}
    
\bibliographystyle{chicago}

\bibliography{reference}
\end{center}
\end{document}